\documentclass[a4paper,fleqn,usenatbib]{mnras}

\usepackage{graphicx}	
\usepackage{amsmath}	
\usepackage{amssymb}	
\usepackage{epstopdf}
\usepackage[para]{threeparttable} 
\usepackage{rotating}
\usepackage{multirow}
\usepackage[usenames, dvipsnames]{color}

\def\kms{\mathrm{km.s}^{-1}}
\def\phidiff{\phi_\mathrm{diff}}

\def\Lsun{\mathrm{L}_{\sun}}
\def\Msun{\mathrm{M}_{\sun}}
\def\My{\mathrm{M}_{\sun}\mathrm{yr}^{-1}}
\def\Tmean{T_\mathrm{eff}}

\def\corate{^{12}\mathrm{CO}/^{13}\mathrm{CO}}
\def\ccrate{^{12}\mathrm{C}/^{13}\mathrm{C}}
\def\vmicro{v_\mathrm{micro}}
\def\vmicromol{v_\mathrm{micro,mol}}
\def\FeH{\mathrm{[Fe/H]}}

\title[CO-multilayer outer atmospheres revealed with VLTI/AMBER]{A CO-multilayer outer atmosphere for eight evolved stars revealed with VLTI/AMBER\thanks{Based on observations performed at the European Southern Observatory, Chile under ESO AMBER Visitor mode program IDs 081.D-0233(A), 092.D-0461(A) and 093.D-0468(A).}\thanks{Observed and reduced data of the eight evolved stars studied in the current paper are available in electronic form at the CDS via: \url{http://vizier.u-strasbg.fr/viz-bin/VizieR?-source=J/MNRAS/489/2595}}}

\author[M. Hadjara et al.]{
M. Hadjara,$^{1,2,3}$\thanks{E-mail: Massinissa.Hadjara@oca.eu}
P. Cruzal\`ebes,$^{4}$
C. Nitschelm,$^5$
X. Chen,$^6$
E. A. Michael,$^{1}$
E. Moreno,$^{1}$
\\
$^{1}$Radio Astronomical Instrumentation Group (RAIG), Terahertz- and Astro-Photonics Laboratory, Departamento de Ingenier\'ia El\'ectrica, \\ Universidad de Chile, Avenida Tupper 2007, Santiago, Chile\\
$^{2}$Instituto de Astronom{\'i}a, Universidad Cat{\'o}lica del Norte, Av. Angamos 0610 Antofagasta, Chile\\
$^{3}$Centre de Recherche en Astronomie, Astrophysique et G\'{e}ophysique (CRAAG), Route de l'Observatoire, B.P. 63, Bouzareah, 16340,\\ Alger, Algeria\\
$^{4}$Universit\'e C\^ote d’Azur (UCA), Centre National de la Recherche Scientifique (CNRS), Observatoire de la C\^ote d’Azur (OCA),\\
 Laboratoire J. L. Lagrange, UMR 7293,Campus Valrose, 06108 Nice Cedex 2, France\\
$^{5}$Centro de Astronom{\'i}a (CITEVA), Universidad de Antofagasta, Avenida Angamos 601, Antofagasta 1270300, Chile\\
$^{6}$Optical Interferometry Group, Shanghai Astronomical Observatory(SHAO), Chinese Academy Sciences(CAS) , Shanghai, 200030, China}

\date{Accepted 2019 August 3. Received 2019 August 3; in original form 2019 May 28}

\pubyear{2019}

\begin{document}
\label{firstpage}
\pagerange{\pageref{firstpage}--\pageref{lastpage}}
\maketitle

\begin{abstract}
We determine the physical parameters of the outer atmosphere of a sample of eight evolved stars, including the Red SuperGiant $\alpha$ Sco, the Red Giant Branch stars $\alpha$ Boo and $\gamma$ Cru, the K giant $\lambda$ Vel, the normal M-giants BK Vir and SW Vir, and the Mira star W Hya (in two different luminosity phases) by spatially resolving the stars in the individual carbon monoxide (CO) first overtone lines. We used the AMBER (Astronomical Multi-BEam combineR) instrument at the Very Large Telescope Interferometer (VLTI), in high-resolution mode ($\lambda/\Delta\lambda\approx12000$) between 2.28 and 2.31 $\mu m$ in K-band. The maximal angular resolution is 10 mas, obtained by triplets telescope configuration, with baselines from 7 to 48 m.
By using a numerical model of a molecular atmosphere in spherical shells (MOLsphere), named \textsc{PAMPERO} (for Physical Approach of Molecular Photospheric Ejection at high-angular-Resolution for evOlved-stars) we add multiple extended CO layers above the photospheric MARCS model at an adequate spatial resolution. We use the differential visibilities and the spectrum to estimate the size ($R$) of the CO MOLsphere, its column density ($N_{\rm CO}$) and temperature ($T_{\rm mol}$) distributions along the stellar radius.
The combining of the $\chi 2$ minimization and a fine grid approach for uncertainty analysis leads to reasonable $N_{\rm CO}$ and $T_{\rm mol}$ distributions along the stellar radius of the MOLsphere.
\end{abstract}


\section{Introduction \label{introduction}}

The Red Giant Branch (RGB) and Asymptotic Giant Branch (AGB) house giant cool evolved stars in late evolutionary stages, particularly of spectral types K and M with low and intermediate-masses $(0.8 \precsim M/\Msun\precsim 8)$.
During their lifetime, the latter can lose up to 30\% of their initial mass \citep{2002A&A...384..452W}, which makes them play a crucial role in the chemical evolution of the galaxy. They enrich the interstellar medium (ISM) by dredging their nuclear-processed material up to the surface before expelling it into the circumstellar environment, while affecting their evolutionary journey at the same time. These features make these stars one of the most important cogwheels of the cosmic mechanism of matter recycling.

According to the amplitudes of their variability, the K and M giant stars can be classified in two principal subtypes:
\begin{itemize}
\item The Mira-type stars, which are in the AGB phase with light amplitudes ranging from 2.5 to 11 in V-magnitude and pronounced periodicity of 80 to 1000 days,
\item and the semi-regular or irregular variable K-M giant stars (simply designated by the term of normal K-M giant stars in this paper) which have amplitudes varying from several hundredths to several magnitudes (usually 1-2 in V-magnitude) and an unclear periodicity.
\end{itemize}
According to the recent list of the General Catalog of Variable Stars (GCVS) 5.1 \citep{2017ARep...61...80S}\footnote{Note that there are many stars in GCVS5 with an unknown spectral type and that \cite{1995A&A...298..159L} found, for more than 40\% of the studied GCVS4 objects, that they are deduced form poor and sparse light curves.} we count 2271 Mira stars and 2235 normal K-M giants (denoted as L, LB, SRA, SRB, SRC and SRD in the catalog) currently clearly identified in the Galaxy.
Even though the numbers are similar, the normal K-M giant stars remain relatively less studied than the Mira stars. Indeed, the Mira-type stars in the AGB phase, which show variability amplitudes much greater than the normal K-M giants, are easier to be detected and studied.
If we focus only on the normal M giants in the AGB phase and compare their mass-loss rates ($\dot{M}$) with those of Mira's stars, we observe that they are nearly the same. Normal K-M giants have $\dot{M}$ between $10^{-8}$ and $10^{-6}\My$ with an expansion velocity ($v_{exp}$) from $3-15\kms$ \citep[e.g.][]{2010A&A...523A..18D,2005AJ....130..842M,2003A&A...411..123G,2003A&A...409..715W,1998ApJS..117..209K}, which means a total mass loss of $0.1\Msun$ during their full AGB lifetime \citep{2014A&A...564A.136O}. For the Mira stars, $\dot{M}=10^{-8}-10^{-6}\My$ with $v_{exp}=3-20\kms$ \citep[e.g.][]{2005AJ....130..842M,2003A&A...409..715W,1998ApJS..117..209K}. It is also important to note that mass-loss-rates of some Miras with substantial dust shells can reach $10^{-5}\My$ and more \citep[][and references therein]{2018A&ARv..26....1H}. Red giant stars near the RGB tip may lose more than $10^{-6}\My$ \citep{2010ApJ...718..522O,2007ApJ...667L..85O,2007PASJ...59S.437I}.
Mass loss chemically feeds the circumstellar environment (CSE) of AGB stars as well as the ISM. The chemical composition of the ejected material is determined by the carbon/oxygen ratio (C/O) \citep[e.g.][]{1973IAUS...52..485W}, from which we can classify those stars into three different types \citep[more details about the evolved stars, their CSE and the most recent works of this field were summarized by][]{2016A&A...588A...4L,2016MNRAS.462..395D,1996A&ARv...7...97H}:
\begin{itemize}
\item Oxygen-rich (O-rich) stars : With C/O$<$1, the manufactured/ejected carbon by the star will quickly bond with oxygen (if temperature and pressure permit) to form very stable carbon monoxide (CO) molecules. The remaining oxygen will form oxygen-rich molecules (such as silicates and oxides) and particles \citep[such as aluminium monoxide, e.g.][]{2017A&A...598A..53D,2007PASJ...59S.437I},
\item Carbon-rich (C-rich) stars : With C/O$>$1, the excess of carbon will produce dust of silicon carbide and graphites \citep{2011AstRv...6h..27G}. Note that \cite{2015MNRAS.446.3277C} observed in their sample of 10 O-rich and 4 C-rich giants more predominant CSE asymmetry for the C-rich stars than for the O-rich stars.
\item S-Type stars : Where C/O$\approx$1, are traditionally considered as in transitional evolutionary step between O-rich and C-rich stars. Note that the abundances of other elements such as lithium and zirconium of O-rich stars are affected by s-process enrichment and hot-bottom burning \citep{2007A&A...462..711G}.
\end{itemize}

The possibility to distinguish an extra-molecular layer from the hot chromosphere and the cool expanding wind was argued the early 1980s based on the CO 2-0 spectra of the Mira stars \citep[even if water vapor spectra were observed on cool luminous stars since early 1960s;][and references therein]{2006ApJ...645.1448T}.
The indisputable evidence for the existence of an extended molecular envelope was first observed, using speckle interferometry technique by \cite{1977ApJ...218L..75L}, on the Miras o Ceti and R Leo, for which they measured angular diameters 2 times larger in the TiO absorption lines than in the continuum. 
Since then, several observations confirmed the presence of these extra-molecular layers around M giants in the AGB phase \citep[][and references therein]{2015A&A...575A..50A,2006ApJ...645.1448T}.
Beside this, more recently, it has been demonstrated that the measured relative size ratio, of these evolved stars, in the continuum and in the absorption band heads of some oxygen-rich molecules (as the silicates and oxides) can reach more than 50$\%$ \citep[][and references therein]{2016A&A...587A..12W,2011A&A...529A.115M}.
Thus, their size appears, at certain wavelengths, much larger than predicted by the classical stellar atmospheric and hydrostatic models (such as MARCS for example).
This extended region of few stellar radii ($R_*$), between the upper photosphere and the innermost part of the circumstellar envelope, where the stellar wind is supposed to take its energy and momentum 
was baptized “MOLsphere” by \cite{2006ApJ...645.1448T}.
The mechanism responsible for mass-loss could be directly related to the physical process that generates the MOLsphere.
The study of the MOLsphere is therefore very important in order to better understand the mass-loss phenomenon of red giant stars in general \citep{1997A&A...320L...1T,2001A&A...376L...1T,2004A&A...418..675P,2004A&A...421.1149O,2007A&A...470..191W,2010A&A...515A..12C,2011A&A...528A.120C,2012A&A...537A..53O,2013A&A...553A...3O,2015A&A...575A..50A,2016A&A...587A..12W,2017A&A...601A...3W,2019A&A...621A...6O}.

Long-baseline spectro-interferometry in the Near-InfraRed (NIR) is an observation tool that allows to resolve and finely study the relatively optically thin and extended MOLspheres of K-M giants.
Indeed, \cite{2012A&A...537A..53O}, by using this technique with the VLTI/AMBER instrument, were able to resolve the carbon monoxide (CO) present in the MOLsphere of a rich panel of K-M giant stars and to determine their sizes, temperature and column density distributions (thanks to a two-layer model), then conclude on the MOLsphere's behavior.
Thus, to extend this work, we decided to study with the same instrument different type of K-M giants, namely : the irregular variable Red SuperGiant (RSG) $\alpha$ Sco, the RGB star with low variability $\alpha$ Boo, the  normal eruptive semi-Variable M-giants $\gamma$ Cru, the normal slow irregular K giant $\lambda$ Vel, the semi-Variable M-giants SW Vir and KB Vir, and the Mira star W Hya at post-minimum light for the observation of February 11$^{th}$ 2014 and at pre-maximum light for the observation of April 22$^{th}$ 2014, with phases of 0.59 and 0.77 respectively \citep[according to][by using light curves of the American Association of Variable Star Observers -AAVSO-]{2016A&A...589A..91O, 2017A&A...597A..20O}.

In this paper, we study and analyze the results obtained by VLTI-AMBER at high spectral resolution in the K-band. We use the numerical model \textsc{PAMPERO} (for Physical Approach of Molecular Photospheric Ejection at high-angular-Resolution for evOlved-stars) in order to constrain the physical proprieties of CO outer layers from the observed flux and visibilities of our target and determine the distributions of the CO column density $N_{\rm CO}$ and the temperature $T_{\rm mol}$ along the stellar radius.
We validate our model by using already published data, namely for $\alpha$ Boo \citep{2018A&A...620A..23O}, BK Vir \citep{2012A&A...537A..53O} and SW Vir \citep{2019A&A...621A...6O}, then for new data, namely $\alpha$ Sco, $\gamma$ Cru, $\lambda$ Vel and W Hya (for two different phase of activity).
In Sec.~\ref{obsdatared}, we present the observations and the data reduction. In Sec.~\ref{Data_int}, we analyze and discuss the final reduced data. In Sec.~\ref{model}, we describe the used model and in Sec.~\ref{Res} we deduce the relevant stellar parameters of our targets and discuss the results. Finally, in Sec.~\ref{conclu} we conclude on the current work.

\section{Observations and data reduction}
\label{obsdatared}

\begin{figure}
\centering
\includegraphics[width=1.\hsize,draft=false]{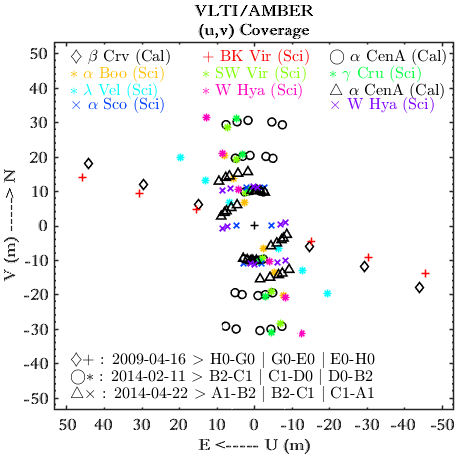}
\caption{Baselines and the corresponding ($u,v$) coverage of VLTI/AMBER observations of our sample of eight K-M giants. Earth-rotation synthesis spanned over $\sim$1.1 h/night. The science targets are represented by colour symbols while the calibrators ($\alpha$ Cen A and $\beta$ Crv) are in black symbols.} \label{All-UV_Coverage}
\end{figure}

Our sample of stars was observed with the AMBER/VLTI instrument \citep{2007A&A...464....1P} with the Auxiliary Telescopes (ATs) triplets B2-C1-D0 (16-32-48m) for 2009 April 16$^{th}$, B2-C1-D0 (10-20-30m) for 2014 February 11$^{th}$ and A1-B2-C1 (7-10-15m) for 2014 April 22$^{th}$. In order to observe the $^{12}$C$^{16}$O (hereafter simply CO) first overtone lines near the 2-0 band head at $2.294\,\mu m$, the observations have been carried out using the high spectral resolution mode of AMBER ($\lambda/\Delta\lambda\approx12000$) in the K-band (HR$\_$K) between $2.28\,\mu m$ and $2.31\,\mu m$. Because of the high brightness of the K-M giants, observed with good weather conditions (0."7-1."2 seeing), the low-contrast fringes (for each of three baselines) were detected without the use of VLTI fringe tracker FINITO \citep{2012SPIE.8445E..1KM}. So, the measurements have been performed with a Detector Integration Time (DIT) of 0.12s and 500 exposures. The sun-like star $\alpha$ Cen A (HD 128620) of spectral type G2.0V \citep{2015ApJ...808..194L} was used as the interferometric and a spectroscopic calibrator for all our science targets, except for BK Vir. For this specific target, the interferometric calibrator $\beta$ Crv could not be used as spectroscopic calibrator. In this case we used the method proposed by \cite{2012A&A...537A..53O} (further explanations below). Table \ref{Tab2} provides the observation log of our sample of targets with the corresponding ($u,v$) coverage shown in Fig.~\ref{All-UV_Coverage}.

\begin{table}
\centering
\caption{VLTI/AMBER observations, with AT Triplets B2-C1-D0 (for the nights of 2009-04-16 and 2014-02-11) and A1-B2-C1 (for the night of 2014-04-22), of our sample of targets with details on the dates, times, and baseline triplets. The calibrators are $\alpha$ Cen A and $\beta$ Crv.} \label{Tab2}
\begin{tabular}{cccc}
\hline \hline
Object & Date \& time & Baseline length & Baseline PA\\
 & & $\textit{B}_{\rm proj}$(m) & \textit{PA}$(^\circ)$\\
\hline \hline
$\beta$ Crv    & 2009-04-16T03:07 & 16,32,48 & 73,73,73 \\
BK Vir         & 2009-04-16T03:47 & 16,32,48 & 68,68,68 \\ 
\hline
$\alpha$ Cen A & 2014-02-11T05:06 & 10,20,30 & 165,165,165 \\
$\gamma$ Cru   & 2014-02-11T05:24 & 10,21,31 & 08,08,08 \\ 
$\alpha$ Cen A & 2014-02-11T05:41 & 10,20,30 & 171,171,171 \\
SW Vir         & 2014-02-11T06:37 & 10,20,29 & 14,14,14 \\
$\alpha$ Cen A & 2014-02-11T06:56 & 10,20,31 & 03,03,03 \\
$\lambda$ Vel  & 2014-02-11T07:13 & 09,19,28 & 45,45,45 \\
$\alpha$ Cen A & 2014-02-11T07:30 & 10,20,30 & 08,09,09 \\
$\alpha$ Boo   & 2014-02-11T07:49 & 07,15,22 & 21,21,21 \\
$\alpha$ Cen A & 2014-02-11T00:17 & 10,20,30 & 14,14,14 \\
W Hya          & 2014-02-11T08:28 & 11,23,34 & 21,21,21 \\
\hline
W Hya          & 2014-04-22T00:38 & 07,11,12 & 03,85,120 \\
$\alpha$ Cen A & 2014-04-22T00:58 & 07,10,16 & 127,73,96 \\
W Hya          & 2014-04-22T01:16 & 08,11,13 & 07,91,129 \\
$\alpha$ Cen A & 2014-04-22T01:34 & 08,10,16 & 140,80,106 \\
$\alpha$ Cen A & 2014-04-22T01:51 & 08,10,16 & 150,86,115 \\
$\alpha$ Cen A & 2014-04-22T01:58 & 09,10,16 & 155,88,119 \\
$\alpha$ Sco   & 2014-04-22T02:13 & 05,11,11 & 179,76,101 \\
$\alpha$ Cen A & 2014-04-22T02:30 & 09,10,16 & 163,94,126 \\
\hline \hline
\end{tabular}
\end{table}

In Fig.~\ref{All-UV_Coverage}, we adopt a color for each star (and for W Hya for each phase). We keep and use this color code in each figure that follows in this current paper.

Thanks to the spectro-interferometric technology of AMBER, we were able to measure, in addition to spectrum, some quantities related to the complex spatial Fourier transform of the brightness of our targets, namely:
\begin{itemize}
\item The differential visibility amplitude which informs us about the shape and the size of the target on several wavelengths between $2.28$ and $2.31\,\mu m$,
\item The differential phase ($\phidiff$) which is related to the photocenter displacement at the first order \citep{2001A&A...377..721J} informs us about the asymmetry as well as the kinematic behavior of the observed upper and outer stellar layers (photosphere, MOLsphere, CSE...etc.),
\item And the closure phase ($\Psi$) which is the sum of all the $\phidiff$ of each baseline. For a point-symmetric objects its value is always equal to zero or $\pi$. $\Psi\neq0$ and $\Psi\neq\pi$ means that the object is asymmetric but the inverse is not always true. Unresolved or partially resolved object has a $\Psi=0$ for example.
\end{itemize}

Our AMBER data have been reduced using version 3.0.9 of the \textit{amdlib}\footnote{Available at \url{http://www.jmmc.fr/data_processing_amber.htm}} software \citep{2009A&A...502..705C, 2007A&A...464...29T}. We adopted a standard frame selection based on fringe (S/N) signal-to-noise \citep{2007A&A...464..107M} and kept the 80\% best-calibrated frames later using the same appropriate reduction methods as those used by \cite{2013A&A...555A..24O,2012A&A...537A..53O,2009A&A...503..183O}, including the error estimation of the calibrated visibilities, $\phidiff$ and $\Psi$. Indeed, thanks to these reduction tools, we first performed a wavelength calibration (with a spectral uncertainty of $1.7\times10^{-5}\,\mu m$ or $\sim 2\,\kms$). We used the identified  calibrator's ($\alpha$ Cen A and $\beta$ Crv) telluric lines from a sample of the atmospheric transmission spectrum, that were measured at Kitt Peak National Observatory\footnote{\url{http://www.eso.org/sci/facilities/paranal/
instruments/isaac/tools/spectra/atmos_S_K.fits}} and were convolved in order to match with the spectral resolution of AMBER's observing mode. Then we converted the wavelength scale to the laboratory frame using heliocentric velocities, of $-24.7\pm0.4\,\kms$ for $\alpha$ CenA and of $-7.6\pm0.1\,\kms$ for $\beta$ Crv, both measured by \cite{2006AstL...32..759G}, without forgetting the sun-earth radial velocity, which is taking account the observation's time and location, from the \textit{IRAF}\footnote{Image Reduction and Analysis Facility, which is available at \url{http://iraf.noao.edu/}} module : \textit{“RVCORRECT”}, in order to convert the observed velocity to the heliocentric frame. For the calibration of our interferometric data, we adopted the uniform-disk diameter of $8.31\pm0.02\,\rm mas$ \citep{2003A&A...404.1087K} for $\alpha$ Cen A\footnote{CHARM2 catalog \citep{2005A&A...431..773R} confirmed the same angular diameter two years later.}. The uniform-disk diameter that we use for $\beta$ Crv is of $3.27\pm0.36\,\rm mas$ \citep[JMMC Stellar Diameters Catalogue - JSDC,][]{2017yCat.2346....0B}, whose the value is close to that of $3.40\pm0.30\,\rm mas$ \citep{2005yCat..34341201R}.
Table \ref{Tab3}, below, summarize all heliocentric velocities (HV) and radial velocities (RV) of all our sample of targets (science and calibrators).

\begin{table}
\begin{minipage}{87mm}
\caption{Heliocentric velocities and radial velocities of our sample of targets (science and calibrators) for each date.} \label{Tab3}
\centering
\begin{threeparttable}
\centering 
\begin{tabular}{ccc}
\hline \hline
Object & Heliocentric Velocity & Radial Velocity\\
 & $(\kms)$ & $(\kms)$\\
\hline \hline
 & Night: 2009-04-16 & \\
\hline
$\beta$ Crv    & $-7.60\pm0.10$$^(\dagger)$ & $-3.83$$^(\ddagger)$ \\
BK Vir         & $16.53\pm0.35$$^(\dagger)$ & $-10.06$$^(\ddagger)$ \\ 
\hline
 & Night: 2014-02-11 & \\
\hline
$\alpha$ Cen A & $-24.7\pm0.4$$^(\dagger)$ & $\sim 22.17$$^(\ddagger)$ \\
$\gamma$ Cru   & $21.0\pm0.6$$^(\dagger)$ & $18.20$$^(\ddagger)$ \\ 
SW Vir         & $-15.0\pm4.4$$^(\dagger)$ & $25.32$$^(\ddagger)$ \\
$\lambda$ Vel  & $17.6\pm0.3$$^(\dagger)$ & $4.60$$^(\ddagger)$ \\

$\alpha$ Boo   & $-5.2\pm0.1$$^(\dagger)$ & $23.13$$^(\ddagger)$ \\
W Hya          & $42.3\pm3.0$$^(\dagger)$ & $25.30$$^(\ddagger)$ \\
\hline
 & Night: 2014-04-22 & \\
\hline
$\alpha$ Cen A & $-24.7\pm0.4$$^(\dagger)$ & $\sim 10.65$$^(\ddagger)$ \\
W Hya          & $42.3\pm3.0$$^(\dagger)$ & $-2.18$$^(\ddagger)$ \\
$\alpha$ Sco   & $-3.5\pm0.8$$^(\dagger)$ & $18.88$$^(\ddagger)$ \\
\hline \hline

\end{tabular}
\begin{tablenotes}
		\footnotesize
		$^(\dagger)$ \citet{2006AstL...32..759G}\\
		$^(\ddagger)$ IRAF (\textit{“RVCORRECT”} module)
\end{tablenotes}
\end{threeparttable}
\end{minipage}
\end{table}

We also use $\alpha$ Cen A for the spectral calibration but, since it is a solar-type star (G2V), we have an excess of CO absorption lines that must be removed. For that purpose we use the method of \cite{2013A&A...555A..24O,2012A&A...537A..53O} and which consists first of calibrating $\alpha$ Cen A's spectrum with the spectrum of the Sun.
The solar flux \citep[observed by][]{1996ApJS..107..312W} is set at the same spectral resolution and wavelength range as our AMBER's data.
In the other hand with $\beta$ Crv, for which no similar spectrum at K-band and in high resolution is available in the literature, we do first \citep[as][]{2012A&A...537A..53O} an auto-spectral-calibration by the theoretical spectrum using the stellar atmosphere model MARCS \citep[][where $\Tmean/\log g/M_\star/\vmicro/\FeH=2800/0.0/0.5/2.0/+0.0$ with moderately CN-cycled composition]{2008A&A...486..951G}, before calibrating the spectrum of BK Vir.

\section{Data interpretation}
\label{Data_int}
This section concern only the new data (i.e. $\gamma$ Cru, $\lambda$ Vel, $\alpha$ Sco and W Hya). For BK Vir, $\alpha$ Boo and SW Vir, data interpretation is well discussed by \cite{2012A&A...537A..53O,2018A&A...620A..23O} and \cite{2019A&A...621A...6O} respectively. For practical reasons, all the figures of this section are gathered in the appendix \ref{Data_int_Fig}.

Figures \ref{gamcru-Data} to \ref{whya2-Data} show the observed visibilities, $\phidiff$, $\Psi$, and spectrum of $\gamma$ Cru (Fig.~\ref{gamcru-Data}), $\lambda$ Vel (Fig.~\ref{lamvel-Data}), $\alpha$ Sco (Fig.~\ref{alpsco-Data}), W Hya for a phase of 0.59 (Fig.~\ref{whya-Data}) and W Hya for an activity phase of 0.77 (Fig.~\ref{whya2-Data}).
The signatures of the CO lines are clearly observable for all stars but even more on W Hya. Despite the short DIT of 0.12s, the lowest visibility that could be measured in the CO lines is $\sim0.1$ for $\gamma$ Cru (Fig.~\ref{gamcru-Data}b), $\sim$ 0.5 for $\lambda$ Vel (Fig.~\ref{lamvel-Data}c), $\sim$ 0.3 for $\alpha$ Sco (Fig.~\ref{alpsco-Data}b and c) and $\sim$ 0.01 for W Hya (Fig.~\ref{whya-Data}c and Fig.~\ref{whya2-Data}c).

Figures \ref{gamcru-Data}d to \ref{whya2-Data}d represent the uniform disk diameters, which were obtained from the respective observed visibilities. They show that the CO first overtone line diameters are 1-17\% larger (25-29 mas) than those in the continuum (24.8 mas) for $\gamma$ Cru (Fig.~\ref{gamcru-Data}d), 5-10\% larger (12-12.5 mas compared to 11.4 mas in the continuum) for $\lambda$ Vel (Fig.~\ref{lamvel-Data}d), 8-10\% larger (39.4-40 mas compared to 36.5 mas) for $\alpha$ Sco (Fig.~\ref{alpsco-Data}d), 21-53\% larger (55-70 mas compared to 46.6 mas) for W Hya (post-minimum phase 0.59, Fig.~\ref{whya-Data}d) and 38-72\% larger (56-70 mas compared to 40.7 mas) for W Hya (pre-maximum phase 0.77, Fig.~\ref{whya2-Data}d). Note that the uniform-disk represent well the object's shape in the continuum, where the star looks bare without MOLsphere (and where its fit is better with a reduced $\chi^2$ between 0-2 for $\gamma$ Cru, 0-0.5 for $\lambda$ Vel and $\alpha$ Sco) while the uniform-disk don’t represent well at all the object's shape of the Mir W Hya in the continuum (with a reduced $\chi^2$  of 0-35 at phase 0.59 and a very high value $\geq$35 at phase 0.77). The uniform-disk approach in the CO lines is worse, because of the detection of the MOLsphere at those lines (where the fit is poor with a reduced $\chi^2$ values between 3-40 for $\gamma$ Cru, 1-7.5 for $\lambda$ Vel, 1-25.7 for $\alpha$ Sco, 5-67 for W Hya at phase 0.59 and $\geq$50 for W Hya 0.77). So the uniform-disk diameter (in the CO lines) can be considered only as a coarse estimation and especially for the Mira star W Hya.

Figures \ref{gamcru-Data}e-h, \ref{lamvel-Data}e-h and \ref{alpsco-Data}e-h show quasi-flat $\phidiff$ and $\Psi$, which means a symmetry in the CO line-forming region, as in the continuum, where the star is detected as naked of any MOLsphere, which means that $\gamma$ Cru, $\lambda$ Vel and $\alpha$ Sco seem to be point-symmetric objects.
Figures \ref{whya-Data}e-h show $\Psi\neq0$ and $\Psi\neq\pi$, which means an asymmetry in the CO line-forming region and that W Hya (with phase of 0.59) seems to be an asymmetric object. The detection of non-zero $\phidiff$ and non-zero/non-$\pi$ $\Psi$ (particularly clear on the 11.3 m baseline, with a peak of 134$^\circ$ at $\lambda=2.294\,\mu m$ and 179$^\circ$ at $\lambda=2.229\,\mu m$), means an asymmetry in the CO line-forming region. Conversely, in the continuum, where the star is detected as naked of any MOLsphere, W Hya (phase of 0.59) seems to be a point-symmetric object.
Same for figures \ref{whya2-Data}e-h, which show $\Psi\neq0$ and $\Psi\neq\pi$ and an asymmetry in the CO line-forming region, which means that W Hya (with phase of 0.77) seems stay an asymmetric object (but more than for phase of 0.59).

By comparing W Hya's uniform disk diameter data (especially in the continuum), which were obtained from the observed visibilities, on both phases, we easily observe that the star's size is significantly bigger on post-minimum phase than on pre-maximum one \citep[as noted by][by comparing two visibilities with the same baseline at different epoch]{2017A&A...597A..20O}.

For $\gamma$ Cru and $\lambda Vel$, when we compare the observed visibilities of representative CO lines to their corresponding spectroscopic line center, we do not observe any shift.  However, for W Hya at the two observed phases, the minima of visibilities switch randomly between blue and red wing of the spectroscopic line. \cite{2011A&A...529A.163O,2009A&A...503..183O} interpreted these asymmetric visibilities as temporally variable inhomogeneous gas on the MOLsphere. According to the obvious wavelength shift of the visibilities with respect to the spectroscopic line center, we suggest that the velocity amplitude of instantaneous inhomogeneous gas for $\gamma$ Cru and $\lambda$ Vel are zero, but that it is not negligible (around $\simeq 25\kms$, which is the velocity resolution of AMBER in high spectral resolution mode) and highly random for W Hya, i.e. MOLspheric velocities of Mira stars which correspond to those found in the literature \citep[e.g.][]{1984ApJS...56....1H,1996A&A...307..481B,2011MNRAS.418..114I}.

\section{Modeling of the AMBER data }
\label{model}

To interpret the AMBER observations of our sample of evolved stars, we used a spectro‐interferometric multilayer MOLsphere model (described below, in Sec.~\ref{pampero}), which surrounds a photosphere with Center-to-Limb intensity Variations (CLV) profiles at each wavelength that we computed by Turbospectrum software \citep{1998A&A...330.1109A,2012ascl.soft05004P}, from MARCS stellar atmosphere models \citep{2008A&A...486..951G} over the observed wavelength range using the CO line list of \cite{1994ApJS...95..535G}.
Each MARCS model\footnote{\url{http://marcs.astro.uu.se}} is specified by some stellar parameters that we should determine beforehand, namely : the effective temperature ($T_{\rm eff}$), surface gravity ($\log g$), micro-turbulent velocity ($v_{\rm micro}$), chemical composition, and stellar mass ($M_\star$). The next subsection precisely describes the determination of these parameters.

\subsection{Determination of stellar parameters}
\label{stellar_param}

Most of all basic stellar parameters of our sample of evolved stars, namely; spectral type, variability type, magnitude, distance, angular diameter ($\diameter_\star$), effective temperature ($\Tmean$), surface gravity ($\log g$), mass ($M_\star$), luminosity ($L_\star$), micro-turbulent velocity ($\vmicro$), metallicity ($\FeH$) and $\corate$ ratio, are available in the literature where they are summarized in Tab.~\ref{Tab4} (appendix \ref{Res_Tab}).
The “References” line of this table (Tab.~\ref{Tab4}) indicates the main references from which parameters values were extracted for each star. Variability types and magnitudes values are taken from GCVS \citep[version 5.1][]{2017ARep...61...80S} for variable stars and from \cite{2002yCat.2237....0D} for what remains. All distances values are from \cite{2007A&A...474..653V} except for W Hydrae which is from \citet{2003A&A...403..993K} \cite[as it was the case for][]{2016A&A...589A..91O,2017A&A...597A..20O}.

\begin{figure}
\centering
\includegraphics[width=1.0\hsize,draft=false]{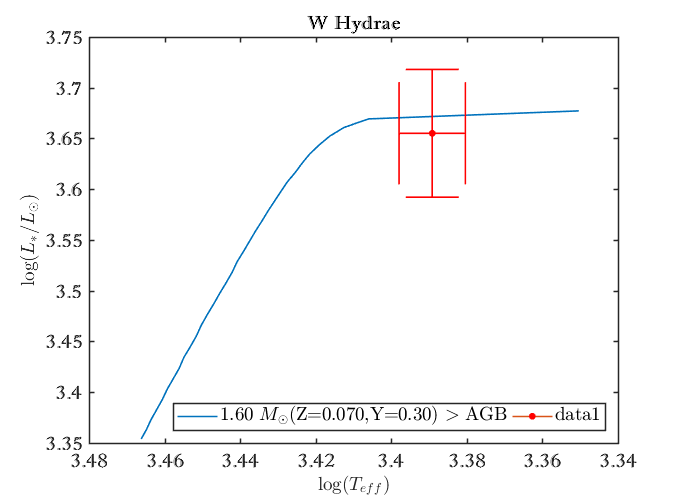}
\caption{H-R diagram with the theoretical evolutionary track of the asymptotic giant branch for a 1.6 $\Msun$ star taken from \citet{2008A&A...484..815B}, with Y=0.30 \& Z=0.07, in blue solid line and the observationally derived positions of the Mira W Hydrae in a red filled circle with error bars.}\label{WHya_Bertelli-AGB}
\end{figure}

We deduce the stellar mass of W Hydrae by comparing its position ($\log(L_\star/\Lsun),\log(T_{\rm eff})$) on the H-R diagram with the theoretical evolutionary tracks of \cite{2008A&A...484..815B}, which contains 912 models of AGB-star with a mass range of $0.55$ to $2.50\,\Msun$.
Figure \ref{WHya_Bertelli-AGB} which compare the observationally derived position of our Mira target together with evolutionary tracks for a 1.2$\,\Msun$ star taken from \cite{2008A&A...484..815B}, suggests that the mass of W Hya is close to 1.6$\,\Msun$ within an uncertainty of $\pm 0.36\,\Msun$ (which corresponds to a surface gravity of $\log g=-0.86\pm0.17$ and a metallicity of $\FeH=0.78$). The parameters that we deduced by our self, namely $M_\star$, $\log g$ and $\FeH$ for W hya are labeled by $^{(\ast)}$ on the Tab.~\ref{Tab4} (appendix \ref{Res_Tab}).

We selected stellar atmosphere models with parameters (detailed in “MARCS model” part  of Tab.~\ref{Tab4}) as close as possible to those available from the spherical MARCS models. The chemical composition (“moderately CN-cycled” or “heavily CN-cycled”) are determined thank to $\corate$ ratio.
Note that although asymmetries in the CO-line-forming region are clearly detected from $\phidiff$ of some stars (as for W Hya for example), we only use a spherical model as a first approximation.

Over each CLV's MARCS model of each star of our sample, we add a MOLsphere thanks to a multilayer model which is described in detail in the next subsection (Sec.~\ref{pampero}).
Using temperature and pressure distributions of the downloaded MARCS model \citep{2008A&A...486..951G} with the cited parameters in Tab.~\ref{Tab4} (appendix \ref{Res_Tab}), we compute the corresponding monochromatic intensity profile and then spectrum \citep[as described in][]{2013A&A...553A...3O}, but this time, using Turbospectrum software \citep{1998A&A...330.1109A,2012ascl.soft05004P}.
In the next subsection (Sec.~\ref{pampero}) we explain further how we use Turbospectrum and introduce our new approach of multilayer MOLsphere modeling.

\subsection{Multi-layer MOLsphere model; \textsc{PAMPERO}}
\label{pampero}
First modeling of a star surrounded by a MOLsphere was done by \cite{2004A&A...418..675P} on FLUOR/IOTA data of Betelgeuse ($\alpha$ Orionis) and Rasalgethi ($\alpha$ Herculis), as black-bodies, with five free parameters, namely: The stellar angular diameter $\diameter_{\star}$, MOLsphere angular diameter (as one layer) $\diameter_{\rm layer}$, stellar temperature $T_{\star}=\Tmean$, MOLsphere temperature $T_{\rm layer}$ and optical depths $\tau_{\rm K}$, $\tau_{\rm L}$, $\tau_{\rm 11.15\mu m}$ at K, L and $11.15\mu m$ bands respectively. 
\cite{2009A&A...503..183O,2011A&A...529A.163O,2012A&A...537A..53O,2013A&A...553A...3O} used a more sophisticated model, thanks to MARCS models, for several different kinds of evolved stars by AMBER/VLTI on K band, where the optical depth is deduced from the column density and temperature of specific molecules (e.g. CO) over one to two MOLsphere layers.
And recently \cite{2014A&A...572A..17M} mixed both approaches using K band AMBER/VLTI data of Betelgeuse for two molecules (CO and H$_2$O) but with the stellar atmosphere grids of ATLAS 9 \citep{2003IAUS..210P.A20C,2005MSAIS...8...14K}.
The approach that we adopt here in this paper is the same that of \cite{2013A&A...553A...3O}, with stellar atmosphere grids of MARCS, using K band AMBER/VLTI data, for the CO molecule, but with continuous multilayer MOLsphere (with a path of $0.1R_{\star}$), in order to study the temperature and molecular density distributions of the different MOLspheres of our sample.

We have titled our model/code \textsc{PAMPERO} for Physical Approach of Molecular Photospheric Ejection at high-angular-Resolution for evOlved-stars. This code, which is written in \texttt{Matlab}\footnote{MATrix LABoratory}, computes first the stellar CLVs thanks to the preselected MARCS models (Tab.~\ref{Tab4}), by using Turbospectrum\footnote{Available here: \url{http://www.pages-perso-bertrand-plez.univ-montp2.fr/}} \citep{1998A&A...330.1109A,2012ascl.soft05004P} and its \textit{SPHLIMB} algorithm, over the observed wavelength range thanks to the listed CO lines of \cite{1994ApJS...95..535G}\footnote{Available with a large list of other molecules here: \url{https://nextcloud.lupm.univ-montp2.fr/s/r8pXijD39YLzw5T}}. These CLVs are computed through the earth atmospheric air. To correct the spectral splitting caused by earth atmospheric air, we use Edl{\'e}n's formula \citep{1966Metro...2...71E} and for the attenuation correction we deduced ourselves from the attenuation factor formula for solar energy \citep{1977STIA...7733445M} for a polluted air at a Zenith angle of $45^{\circ}$ as average approximation.
The monochromatic visibilities are deduced by a simple Fourier transform of the monochromatic intensity profiles, then all the MARCS outputs were convolved with the AMBER’s spectral resolution ($\lambda/\Delta\lambda=12000$), namely the intensity profile, visibility, and spectrum.
We adopt values of the modeled angular diameter which make the optimum correspondence between the modeled and observed visibilities in the continuum for the three observational baselines.
For all our sample of evolved stars, although our AMBER data spatially resolve the MOLsphere in the individual CO lines and the modeled MARCS atmosphere alone predicts well the spectra, this is not sufficient to explain the observed visibilities behavior. Indeed, the predicted visibilities in the CO lines by the MARCS model are too high, which means that either the extension of the real CO-line-forming layer is much higher than the MARCS model prediction or there is an additional component contributing to the CO lines above the MARCS photosphere modeling. So, and as can be seen, the theoretical spectra of strong molecular or atomic features, which are deduced from only the photospheric models can be highly misleading. This was well demonstrated, in the past, for BK Vir, $\alpha$ Boo and SW Vir by \cite{2012A&A...537A..53O,2018A&A...620A..23O} and \cite{2019A&A...621A...6O} respectively, by using a MOLsphere model of two-layers.

\begin{figure}
\centering
\includegraphics[width=1.0\hsize,draft=false]{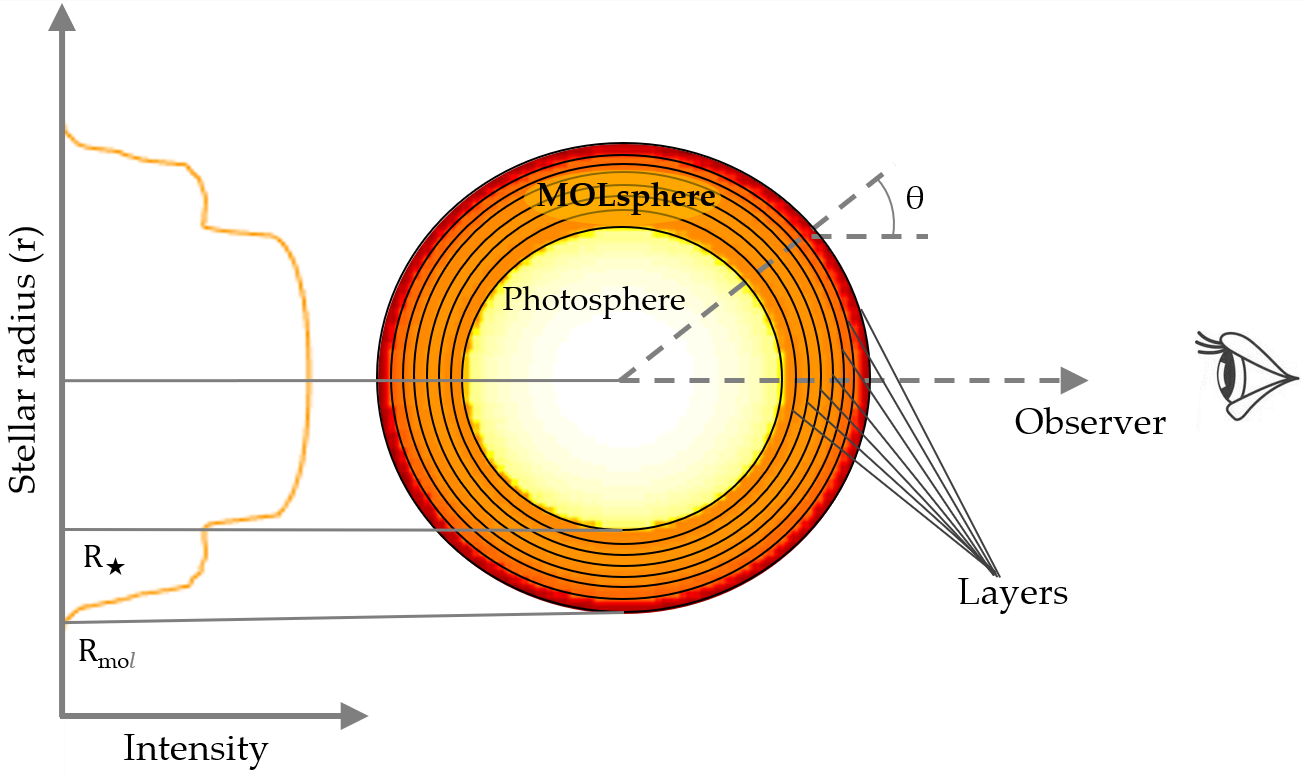}
\caption{Schematic view of \textsc{PAMPERO}, where a continuous multilayer MOLSphere is surrounding a CLV MARCS model.}\label{PAMPERO_Schema}
\end{figure}

So, we need to add MOLspheres over our MARCS model CLVs to get visibilities which will be more consistent with the observations. Figure~\ref{PAMPERO_Schema} depicts a schematics view of our best model for SW Vir (as for example, see Sec.~\ref{Res}), at $\lambda=2.2936\mu m$, with an angular diameter $\diameter_\star$ and which is surrounded by a continuous multilayer MOLsphere.
The Eq.~\ref{eq1} depicts the analytic expression $I_{\rm \star+mol}(\diameter_\star,\mu,\lambda,T_{\rm mol},N_{\rm mol})$ of the CLV star and MOLsphere and which we will call simply $I$, where $\lambda$ is the wavelength, $T_{\rm mol}$ temperature molecular distribution, $N_{\rm mol}$ column density molecular distribution and $\mu=\sqrt{1-(r/R_{mol})^2}=\cos\theta$ that is related to the stellar radius $r$ (which varies from $0$ to the MOLsphere outer radius $R_{\rm mol}$ passing through the star radius $R_\star$) and $\theta$ the angle between star center direction and the line of sight, as follow \citep[e.g.,][]{2014A&A...572A..17M}:
\begin{equation}\label{eq1} 
I=\left\{
    \begin{array}{ll}
        I_\star\exp\left[-\mu\tau_{\rm mol}\right]+I_{\rm mol}(1-\exp\left[-\mu\tau_{\rm mol}\right]) & \mbox{if } \mu\leq \mu_\star \\
				& \\
        I_{\rm mol}(1-\exp\left[-2\mu\tau_{\rm mol}\right]) & \mbox{if } \mu>\mu_\star
    \end{array}
\right.
\end{equation}
Where $\mu_\star=\sqrt{1-(R_\star/R_{\rm mol})^2}$ and $\tau_{\rm mol}$ is the optical depth of the MOLsphere. $I_{\rm mol}$ denotes according to $(\lambda,T_{\rm mol})$, the Planck (black body) function of the MOLsphere, given by:
\begin{equation}\label{eq2} 
I_{\rm mol}=\frac{2hc^2}{\lambda^5}\frac{1}{\exp(hc/\lambda K_{\rm B}T_{\rm mol})-1},
\end{equation}
where $T_{\rm mol}$ is the temperature of the MOLsphere, $h$ the Planck constant, $c$ light velocity, $K_{\rm B}$ the Boltzmann constant and $I_\star$ is the MARCS monochromatic stellar intensity according to $(\lambda,\mu)$, defined as follow \citep[e.g.,][]{2013A&A...553A...3O}:
\begin{equation}\label{eq3}
I_\star=\int S_\lambda(\tau_\star)\exp(-\tau_\star)d\tau_\star,
\end{equation}
where $S_\lambda$ is the source function and $\tau_\star$ the stellar monochromatic optical depth at $\lambda$ \citep[for further details see][]{2008A&A...486..951G}.
For the MOLsphere, the monochromatic $\tau_{\rm mol}$ is related to $\lambda$, the temperature and column density molecular distributions along the stellar radius $T_{\rm mol}(r/R_\star)$ and $N_{\rm mol}(r/R_\star)$ (that we define for more simplicity as $T_{\rm mol}$ and $N_{\rm mol}$ respectively) and which is given by:
\begin{equation}\label{eq4}
\begin{array}{l}
\tau_{\rm mol}=\sum_{r/R_\star} N_{\rm mol}\sigma_{\rm eff}\frac{gf}{Q(T_{\rm mol})}[1-\exp(\frac{hc}{\lambda K_{\rm SB}T_{\rm mol}})],
\end{array}
\end{equation}
where $K_{\rm SB}$ define the Stefan-Boltzmann constant, $\sigma_{\rm eff}$ the cross-section, $Q(T_{\rm mol})$ the partition function specific for each molecule and $gf$-value is the absorption oscillator strength \citep[for further details see][]{1994ApJS...95..535G,2015PASP..127..266M}.

If we can deduce the spectrum directly from $I$ (Eq.~\ref{eq1}) by this simple and usual formula \citep[e.g.,][]{2013A&A...553A...3O}:
\begin{equation}\label{eq5}
F=2\pi\int_0^1 I(\mu) \mu d\mu,
\end{equation}
we have to derive the interferometric observables from the monochromatic intensity distribution of our target ($I_{\rm \star+mol}$; Eq.~\ref{eq1}) thanks to the Van Cittert-Zernike theorem \citep{1934Phy.....1..201V,1938Phy.....5..785Z}. Because of the spherical symmetry specificity of cool evolved stars, it is more practical to derive the visibility $V(B,\lambda)$ using the Hankel transform and utilizing the assumed $\diameter_\star$ of the studied object for each projected baseline $B$ at $\lambda$, as follow \citep[e.g.,][]{2014A&A...572A..17M}:
\begin{equation}\label{eq6}
V=\frac{2\pi\int_0^{R_{\rm mol}} I J_0(\pi B\diameter_\star r/\lambda)rdr}{F},
\end{equation}
where $I$ is the CLV of a star surrounded by MOLsphere as defined in Eq.~\ref{eq1}, $J_0$ zeroth order Bessel function of the first kind, and $r$ the radius along the star and its MOLsphere (see Fig.~\ref{PAMPERO_Schema}).

So, from the visibility (Eq.~\ref{eq6}; $V=\left|V\right|e^{-i\phidiff}$) we deduce the differential phase $\phidiff(\lambda,B)$ and the closure phase $\Psi(\lambda)=\sum_B \phidiff(\lambda,B)$.
As explained previously; our model is spherically symmetric, which means that all the $\phidiff$ on the first visibility lobe are zero, whereas $\phidiff$ sign is flipped in the second visibility lobe with a $\pi$ value.
At the end and for more rigorous comparison between the model and the observations, we interpolate our modeled spectro-interferometric data $(F,V,\phidiff\;\&\;\Psi)$ at observed wavelengths $(\lambda_{obs})$ which were converted to the laboratory frame previously (as explained in the Sec.~\ref{obsdatared}).

Regarding the temperature and density distributions along the stellar radius $(r/R_\star)$, namely the molecular column density (in this paper we focused only on the carbon monoxide molecule $N_{\rm mol}=N_{\rm CO}$) and the MOLsphere's temperature $T_{\rm mol}$ of $I_{\rm \star+mol}(\mu,\lambda,T_{\rm mol},N_{\rm mol})$ of Eq.~\ref{eq1}, we assume that they are inversely proportional to $(r/R_\star)^{\zeta}$, as follow:
\begin{equation}\label{eq7} 
\left\{
    \begin{array}{l}
			  T_{\rm mol}=T_0\big[\big(\frac{r}{R_{\rm mol,0}}\big)^{-\zeta_{\rm T}}\big],\\
        N_{\rm mol}=N_0\big[\big(\frac{r}{R_{\rm mol,0}}\big)^{-\zeta_{\rm N}}\big],
    \end{array}
\right.
\end{equation}
where $\zeta_{\rm T}$ and $\zeta_{\rm N}$ are coefficients, $T_0$ and $N_0$ the temperature and the column density of the first MOLsphere layer respectively, and $\frac{r}{R_{\rm mol,0}}$ derived from the normalization $\frac{r}{R_\star}/\frac{R_{\rm mol,0}}{R_\star}$, where $R_{\rm mol,0}$ is the inner radius of the first MOLspheric layer.

No matter the number of the MOLsphere layers, by fixing the thickness (as $0.1R_\star$ for example), our numerical model needs only six free parameters, namely: $T_0$, $N_0$, $\zeta_{\rm T}$, $\zeta_{\rm N}$, $R_{\rm mol,0}$ and $R_{\rm mol,end}$, where $R_{\rm mol,0}$ and $R_{\rm mol,end}$ are the upper and lower radius boundaries of our MOLsphere.
We adjust also two other parameters; the angular diameter $\diameter_\star$ in order to settle modeled visibilities at the same level of the observed ones at the continuum, and the MOLspheric micro-turbulent velocity ($\vmicromol$) which affects only the width of the molecular lines. In general, we find $\vmicromol$ values in the literature. Otherwise, we got the best model with the six free parameters cited above ($T_0,\,N_0,\,\zeta_{\rm T},\,\zeta_{\rm N},\,R_{\rm mol,0},\,\&\,R_{\rm mol,end}$), we adjust $\diameter_\star$ within its bounds of uncertainty and we play with $\vmicromol$ until we get the best match, in terms of the width of the molecular lines, with observations (with the minimum value of $\chi^2$).
 
To avoid an overlap with the MARCS photospheric model, which is extended up to $1\,R_\star$ for all our sample of stars, the radius of the first/inner layers of all our MOLsphere models is set to be equal to, or larger than the maximum size of the photospheric MARCS model of each studied star. Table.~\ref{Tab5} summarizes the maximum size of all MARCS photospheric models of our targets, star by star.

\begin{table}
\centering
\caption{Table summarizing the maximum size of photospheric MARCS model (Max($R_{\rm marcs}$) in function of $r/R_\star$) of each studied star of our sample.} \label{Tab5}
\begin{tabular}{cc}
\hline \hline
Star & Max($R_{\rm marcs}$) \\
\hline
BK Vir & 1.1581$R_\star$ \\ 
$\alpha$ Boo & 1.0207$R_\star$ \\
SW Vir & 1.1581$R_\star$ \\
$\gamma$ Cru & 1.0358$R_\star$ \\
\hline \hline
\end{tabular}
\begin{tabular}{cc}
\hline \hline
Star & Max($R_{\rm marcs}$) \\
\hline
$\lambda$ Vel & 1.0159$R_\star$ \\
$\alpha$ Sco & 1.0541$R_\star$ \\
W Hya & 1.1099$R_\star$ \\
 & \\
\hline \hline
\end{tabular}
\end{table}

\begin{figure*}
\centering
\includegraphics[width=1.0\hsize,draft=false]{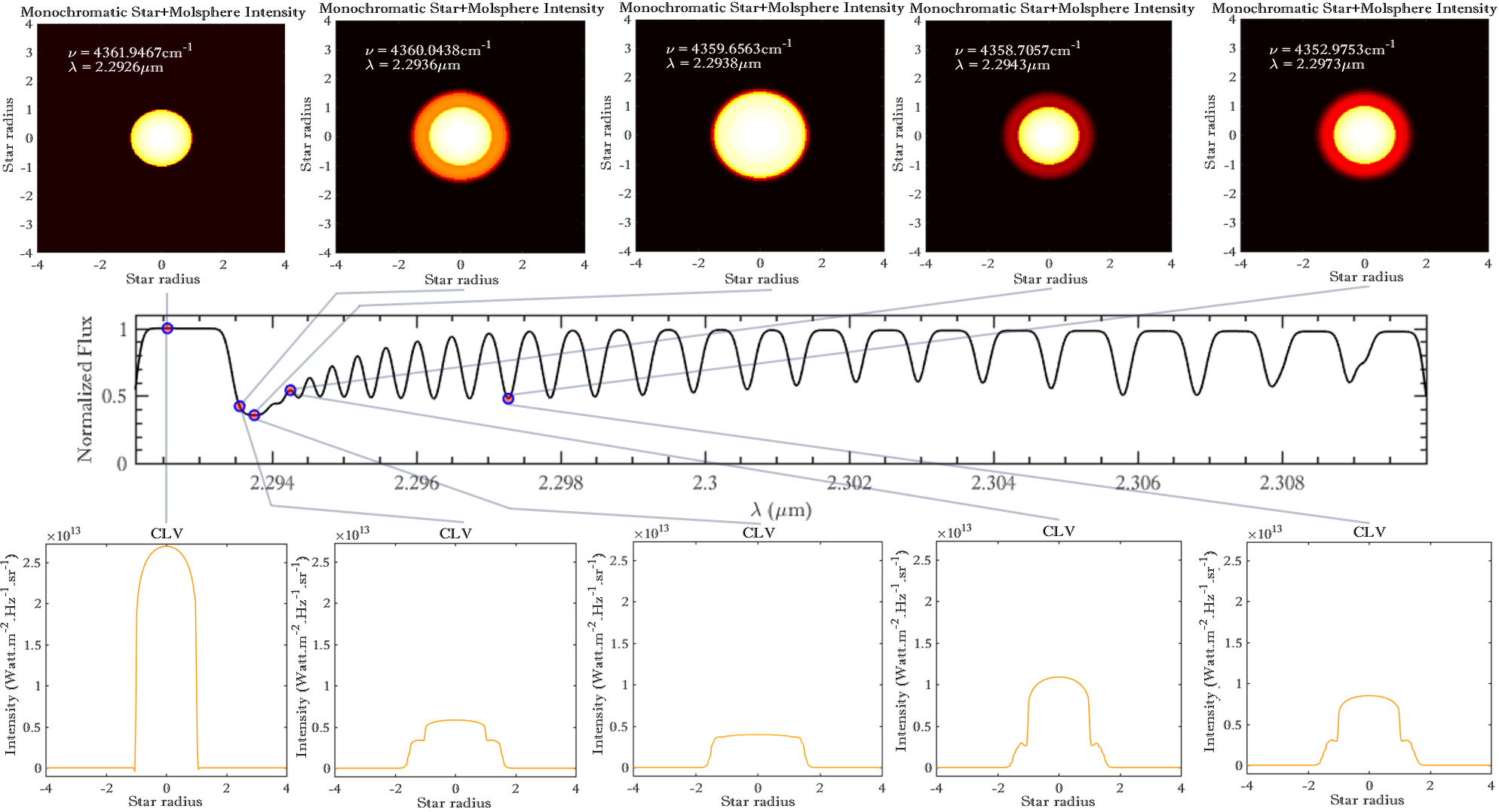}
\caption{\textsc{PAMPERO}'s intensity maps and CLVs, according the normalized flux, for the best model for SW Vir, at the continuum ($\lambda=2.2926\mu m$), around the CO band head ($\lambda=2.2936\;,2.2938\;\&\;2.2943\mu m$) and on one individual CO line ($\lambda=2.2973\mu m$).}\label{SWVir-Flux-Int-lambda}
\end{figure*}

Figure \ref{SWVir-Flux-Int-lambda} describes intensity maps and CLVs of our code \textsc{PAMPERO}, according to the normalized flux, for the best model for SW Vir (see Sec.~\ref{Res}), at 5 different wavelengths; at the continuum ($\lambda=2.2926\mu m$), around the CO band head ($\lambda=2.2936\;,2.2938\;\&\;2.2943\mu m$) and on one individual CO line ($\lambda=2.2973\mu m$).

In the next section (Sec.~\ref{Res}), and by combining a $\chi^2$ minimization method and a fine grid approach, we summarize according to the observations, the best modeled molecular behavior that we found thanks to \textsc{PAMPERO} ($T_{\rm mol}$ and $N_{\rm co}$), target by target.

\section{Results and discussions}
\label{Res}

We present the best \textsc{PAMPERO} models that were found for the entire sample of targets. Indeed, by using a new atmospheric temperature-density distribution approach we were able to resolve the extended MOLspheres for several kinds of evolved stars, including normal K-M giants, Mira and red giants, in the IR (K band) for CO, as shown in Fig.\,\ref{BKVir-Res} to \ref{whya2-Res}. For practical reasons, all the results of this section are summarized in the Table\,\ref{Tab4} and the corresponding figures are shown in the Appendix\,\ref{Res_Fig}. The uncertainties calculated by our $\chi^2$ minimization method, are described in \citet[Section\,4][]{2014A&A...569A..45H}.

We present our results as follow: We follow the chronological order of observations, and maintain the same colour code as defined in Section.~\ref{obsdatared} and presented in Fig.~\ref{All-UV_Coverage}. For the best model of each star, we show first the MOLsphere's carbon monoxide column density $N_{\rm CO}$ and temperature $T_{\rm mol}$ along the stellar radius $(r/R_\star)$. Next, we present the spectro-interferometric data, namely the normalized flux $(F/F_{\rm Continuum})$, the visibilities ($V$) corresponding to their respective baselines ($B$) and closure phase $\Psi$, all according to the wavelength $(\lambda)$, overplotted with the results of our best model (continuous black line). In the same figure, we show in the title, next to the name of the star, the reduced $\chi^2$-value (hereafter simply $\chi^2$).
By applying our new continuous and multilayer MOLsphere approach, we present first in this section the results of three published stars, BK Vir, $\alpha$ Boo, and SW Vir (where we compare, for these latter, the results of our multilayer approach with the results from the bi-layers approach, using the values given in Table.~\ref{Tab4} and computations with \textsc{PAMPERO}), then we present the new results of four evolved stars (including one Mira observed for two epochs), $\gamma$ Cru, $\lambda$ Vel, $\alpha$ Sco and W Hya (Fig.\ref{BKVir-Res} to \ref{whya2-Res}).
We determine the uncertainties of the six free parameters ($T_0,\,N_0,\,\zeta_{\rm T},\,\zeta_{\rm N},\,R_{\rm mol,0},\,\&\,R_{\rm mol,end}$) that we presented in Section\,~\ref{pampero}, using the same method that was used for the other model, which is dedicated to Hot Active Stars (SCIROCCO; \citealt{2014A&A...569A..45H, Massi2015,2018MNRAS.480.1263H}), when the $\chi^2$-minimization was used, after localizing the global minimums by using a large grid of six free parameters. The results are given in Tab.~\ref{Tab6} in Appendix \ref{Grid_Tab}. For $\alpha$\,Sco we improved the fitting with an ``ad hoc'' manipulation; by manually adjusting the values $T_{\rm mol}$ and $N_{\rm co}$ for a few layers, by using a priori that we determined from the MOLspheric model for this star. We assumed that $\alpha$\,Sco has a MOLsphere with a convective behavior \citep{2017Natur.548..310O}.

\subsection{BK Virginis}
\label{BKVir}
The best \textsc{PAMPERO} model fit for BK Vir found a total thickness of $3.3R_\star$ containing 33 layers, each with a thickness of $0.1R_\star$. We found a temperature-density distribution of $T_{\rm mol}(r/R_\star)$ and $N_{\rm CO}(r/R_\star)$, according to Eq.~\ref{eq7}, with a MOLspheric temperature of the first layer of $T_0=2010K$ with a coefficient of $\zeta_{\rm T}=0.35$, and a MOLspheric CO column density of the first layer at $N_0=10^{22.3}mol/cm^{-2}$ with a coefficient of $\zeta_{\rm N}=18$.
The first two plots in Fig.\,\ref{BKVir-Res} depict the temperature-density distribution $T_{\rm mol}(r/R_\star)$ and $N_{\rm CO}(r/R_\star)$ respectively, where both distributions start from $R_{\rm mol,0}=1.2R_\star$ and decrease to $R_{\rm mol,end}=4.5R_\star$ with the coefficients of $\zeta_{\rm T}$ and $\zeta_{\rm N}$ respectively.
The MOLspheric temperature-density distribution shows a decrease in temperature and $N_{\rm CO}$ from $T_{\rm mol}\big(\frac{R_{\rm mol,0}}{R_\star}\big)=2010K$ to $T_{\rm mol}\big(\frac{R_{\rm mol,end}}{R_\star}\big)=1266K$ and $N_{\rm CO}\big(\frac{R_{\rm mol,0}}{R_\star}\big)=10^{22.3}$ to $N_{\rm CO}\big(\frac{R_{\rm mol,end}}{R_\star}\big)=10^{12}mol/cm^{-2}$ respectively.
We adjusted $\diameter_\star=10.6mas$ to obtain the best fit between the model and the the observed visibilities in the continuum, and according to \citet[][and references therein]{2012A&A...537A..53O}, we fixed $\vmicromol=\vmicro=4\kms$ in order to get the best agreement between the width of modeled molecular lines and the observations.
The bottom portion of Fig.\,\ref{BKVir-Res} shows a comparison between the modeled and observed normalized flux $F/F_{\rm Continuum}(\lambda)$, visibilities $V(\lambda)$ and closure phases.
					
By examining the visibility plots, we observe slight differences in the variations of the modeled data with respect to observed ones at the continuum (Fig.~\ref{BKVir-Res}), which means that the target is not perfectly spherical.
\cite{Miguel2014} demonstrated that hot or dark spot(s) could affect visibility. We interpret the discrepancy in the fit of the CO band head for the second visibility ($B=31.82m$) due to the presence of a hot or a dark spot(s) on the edge of the MOLsphere (around $4.4R_\star$), which only appears around $\lambda=2.294\mu m$ (e.g., Fig.~\ref{SWVir-Flux-Int-lambda}). We cannot determine the position or the number of spots with our sparse $(u,v)$ coverage (Fig.~\ref{All-UV_Coverage}).
Alternatively, the discrepancy between the model and observed spectrum at $\lambda=2.299\,\&\,2.305\mu m$ is due to the residuals of the strong telluric lines, which appear in all the observed spectra of all targets.

At first glance, the best continuous multilayer MOLsphere model (\textsc{PAMPERO}) for BK Vir agrees with that of the discontinuous two-layers presented by \citet{2012A&A...537A..53O}, specially at the first layer, where \citet{2012A&A...537A..53O} found $1.2-1.25R_\star$ $(T_0=1900-2100K,\,N_0=1-2\times10^{22}\,mol/cm^{-2})$ while, our results show $1.18-1.22R_\star$ $(T_0=1960-2060K,\,N_0=1.6-2.5\times10^{22}\,mol/cm^{-2})$. However, we find noticeable disagreement for the second layer, especially for $N_{\rm CO}$ where \citet{2012A&A...537A..53O} found $2.5-3R_\star$ $(T=1500-2100K,\,N=10^{19}-10^{20}\,mol/cm^{-2})$ and our results indicate $(T=1400-1600K,\,N=10^{15}-10^{17}\,mol/cm^{-2}$, for the same size). We attribute this large difference in density to the fact that the discontinuous two-layers do not take into account the effect of a continuous MOLsphere which may overestimate the temperature-density parameters of a large MOLsphere area when using a single thin layer.

Note that we also checked the results of BK Vir by another UD value of its calibrator $\beta$ Crv ($3.40\pm0.30\,\rm mas$; \citealt{2005yCat..34341201R}) instead of JSDC ($3.27\pm0.36\,\rm mas$; \citealt{2017yCat.2346....0B}). There is no significant difference between both results, except the adjusted $\diameter_\star$ of BK vir which is to $10.75mas$ instead of $10.6mas$.
			
\subsection{$\alpha$ Boo}
\label{alpBoo}
For $\alpha$ Boo, the best \textsc{PAMPERO} model fit shows a total thickness of $0.5R_\star$ for five layers each with a thickness of $0.1R_\star$. The temperature-density distribution is of $T_0=1650K$ with a coefficient of $\zeta_{\rm T}=6.56$, and $N_0=10^{19.2}mol/cm^{-2}$ with a coefficient of $\zeta_{\rm N}=30$.
The first two plots in Fig.\,\ref{alpBoo-Res} depict $T_{\rm mol}(r/R_\star)$ and $N_{\rm CO}(r/R_\star)$, where both distributions start from $R_{\rm mol,0}=2.5R_\star$ and decrease to $R_{\rm mol,end}=3R_\star$ with the coefficients of $\zeta_{\rm N}$ and $\zeta_{\rm T}$ respectively. 

The MOLspheric temperature-density distribution shows a decrease in temperature and $N_{\rm CO}$ from $T_{\rm mol}\big(\frac{R_{\rm mol,0}}{R_\star}\big)=1650K$ to $T_{\rm mol}\big(\frac{R_{\rm mol,end}}{R_\star}\big)=499K$ and $N_{\rm CO}\big(\frac{R_{\rm mol,0}}{R_\star}\big)=10^{19.2}mol/cm^{-2}$ to $N_{\rm CO}\big(\frac{R_{\rm mol,end}}{R_\star}\big)=10^{16.8}mol/cm^{-2}$ respectively.
We adjusted $\diameter_\star=20.7mas$ to obtain the best fit between the model and the observed visibilities in the continuum, whilst we fixed $\vmicromol=\vmicro=2\kms$  \citep[as discussed by][and references therein]{2014A&A...561A..47O}, in order to get the best agreement between the width of modeled molecular lines and the observations.
The bottom portion of Fig.\,\ref{alpBoo-Res} shows a comparison between modeled and observed normalized flux $F/F_{\rm Continuum}(\lambda)$, visibilities $V(\lambda)$ and closure phases.		
					
We observe on visibilities slight differences in the variations of modeled data with respect to observed ones at the continuum (Fig.~\ref{alpBoo-Res}), which means that the target is not perfectly spherical. On the third visibility, at the third baseline $(B=21.83m)$, our fit has a lower goodness of fit with respect to the first visibilities ($B=7.27m$ and $B=14.56m$) but it remains inside the 1-$\sigma$ uncertainty, which is acceptable. 

By assuming a continuity of the MOLsphere, we are with this work adding another scenario to the case of $\alpha$ Boo's MOLsphere. While \citet{2018A&A...620A..23O} proposes two different MOLspheric layers, where the inner layer is close to the photosphere at $1.04\pm 0.02R_\star$ with $T_{\rm mol}=1600\pm 400K$ \& $N_{\rm CO}=10^{20\pm 0.3}mol/cm^{-2}$, and the outer one is at $2.6\pm 0.2R_\star$ with $T_{\rm mol}=1800\pm 100K$ \& $N_{\rm CO}=10^{19\pm 0.15}mol/cm^{-2}$, we suggest a possibility of continuous distribution. Indeed, \citet{2003ApJ...598..610A} argues that RGB stars host a cool CO area (which they called COMosphere $\sim 1000K$) between the photosphere and the chromosphere, and which is crossed by an important stellar wind because of the intense magnetic field of this kind of stars \citep[Alfv\'{e}n-wave-driven wind, e.g.,][]{2007ApJ...659.1592S,2010ApJ...723.1210A}.
Our best \textsc{PAMPERO} model proposes a continuous distribution with $T_{\rm mol}$-value close to that of the inner layer given by \citet{2018A&A...620A..23O} while a $N_{\rm co}$-value close to the outer value given by \citet{2018A&A...620A..23O}, with a lower temperature of $\sim1300\,K$ at $2.6R_\star$. The magnetic field loops combined with the stellar wind, which expels the photospheric matter over long distances, where we suspect that the COMosphere may overlap the chromosphere (which remains a topic of debate) could explain the heating mechanism which is present in the outer atmosphere.
Maybe Arcturus has a discontinued MOLspheric distribution as suggested by \cite{2018A&A...620A..23O}, but our model demonstrates that the only continuous MOLspheric distribution which is possible, is that from $2.5\pm0.2R_\star$, where we checked every conceivable scenario of continuous MOLspheric distributions.

\subsection{SW Vir}
\label{swvir}
Our best \textsc{PAMPERO} model fit for SW Vir shows a total thickness of $1.8R_\star$ for 18 layers each with a thickness of $0.1R_\star$. The temperature-density distribution is of $T_{\rm mol}(r/R_\star)$ and $N_{\rm CO}(r/R_\star)$, according to Eq.~\ref{eq7}, has a MOLspheric first layer temperature of $T_0=1950K$ with a coefficient of $\zeta_{\rm T}=0.06$, and a MOLspheric CO column density $N_0=10^{22.5}mol/cm^{-2}$ with a coefficient of $\zeta_{\rm N}=30$.
The first two plots in Fig.\,\ref{SWVir-Res} depict the temperature-density distribution $T_{\rm mol}(r/R_\star)$ and $N_{\rm CO}(r/R_\star)$ respectively, where both distributions start from $R_{\rm mol,0}=1.2R_\star$ and decrease to $R_{\rm mol,end}=3R_\star$ with the coefficients of $\zeta_{\rm T}$ and $\zeta_{\rm N}$ respectively.
The MOLspheric temperature-density distribution shows a decrease in temperature and $N_{\rm CO}$ from $T_{\rm mol}\big(\frac{R_{\rm mol,0}}{R_\star}\big)=1950K$ to $T_{\rm mol}\big(\frac{R_{\rm mol,end}}{R_\star}\big)=1846K$ and from $N_{\rm CO}\big(\frac{R_{\rm mol,0}}{R_\star}\big)=10^{22.5}$ to $N_{\rm CO}\big(\frac{R_{\rm mol,end}}{R_\star}\big)=10^{10.6}$ respectively.
We adjusted $\diameter_\star=16.7mas$ to obtain the best fit between the model and the observed visibilities in the continuum, and according to \citet[][and references therein]{2019A&A...621A...6O} we fixed $\vmicromol=\vmicro=3.6\kms$ in order to get the best agreement between the width of modeled molecular lines and the observations.
The bottom portion of Fig.\,\ref{SWVir-Res} shows a comparison between the modeled and observed normalized flux $F/F_{\rm Continuum}(\lambda)$, visibilities $V(\lambda)$ and closure phases.							
					
From the visibilities, we can observe that our modeled data are of approximately the same amplitudes with respect to the observations at the continuum (Fig.~\ref{SWVir-Res}), which corresponds to a quasi-perfectly spherical target.
As for BK Vir, we interpret the discrepancy in the fit on the CO band head (and some other individual lines) of the second visibility ($B=19.64m$) due to a presence of a hot or a dark spot(s) on the MOLsphere and which we cannot determine the position nor the number with our sparse $(u,v)$ coverage \citep[as demonstrated by][]{Miguel2014}.

As for BK Vir, at first glance, our best continuous multilayer MOLsphere model (\textsc{PAMPERO}) for SW Vir agree within their 1-$\sigma$ uncertainties with that of discontinuous two-layer solution given by \citet{2019A&A...621A...6O}, particularly at the first layer, where \citet{2019A&A...621A...6O} found  $1.3\pm 0.1R_\star$ $(T_0=2000\pm 100K,\,N_0=1-3\times10^{22}\,mol/cm^{-2})$ while our results show at the same size $(T_0=1940\pm 60K,\,N_0=10^{22.5}-10^{20.5}\,mol/cm^{-2})$. However, we find noticeable disagreement for the second layer, especially for $N_{\rm CO}$ where \cite{2019A&A...621A...6O} found $2.0\pm 0.2R_\star$ $(T=1700\pm 100K,\,N=2\times 10^{19}-2\times 10^{20}\,mol/cm^{-2})$ while our results indicate at the same size $(T=1880-1903K,\,N=10^{14.6}-10^{17.22}\,mol/cm^{-2})$. We attribute this large difference in density, for BK Vir, to the fact that the discontinuous two-layers do not take into account the effect of a continuous MOLsphere which may overestimate the temperature-density parameters of a large MOLsphere area when using a single thin layer.

Axial symmetry is not usual for normal K-M giant stars. We observe axisymmetrical CSEs, in general, on post-AGB stars and planetary nebulae \citep{2006ApJ...651..288D}, whereas most AGB stars shows spherically symmetric envelopes \citep[especially O-rich stars][]{2015MNRAS.446.3277C}. \citet{2009A&A...498..627F} suspected, thanks to a Hipparcos radial velocity study of a sample of M giants, that SW Vir can be a binary system. \citet{2011AstRv...6g..27N} used Gemini/MICHELLE IR data of SW Vir and deduced recently, from the spectra, an axisymmetric dust shell which can not be attributed to a simple radial temperature variation. Hence, they strongly favor the influence of a companion.
Unfortunately, because of the array configuration of our sparse ($u,v$) coverage (see Fig.~\ref{All-UV_Coverage}) it is impossible for us to deduce, by projection, the corresponding declination and right ascension photocenter displacements and therefore we are unable to confirm the existence of any axial symmetry of the CSE of SW Vir as observed by \cite{2011AstRv...6g..27N}.

\subsection{$\gamma$ Cru}
\label{gamcru}
The best \textsc{PAMPERO} model fit for $\gamma$ Cru found a total thickness of $0.1R_\star$ containing a single layer. We found a temperature-density distribution of $T_0=960K$ with a coefficient of $\zeta_{\rm T}=0.1$, and $N_0=10^{21.5}mol/cm^{-2}$ with a coefficient of $\zeta_{\rm N}=50$.
The first two plots in Fig.\,\ref{gamCru-Res} depict the temperature-density distribution $T_{\rm mol}(r/R_\star)$ and $N_{\rm CO}(r/R_\star)$, where both distributions start from $R_{\rm mol,0}=5.15R_\star$ and slightly decrease until $R_{\rm mol,end}=5.25R_\star$ with the coefficients of $\zeta_{\rm T}$ and $\zeta_{\rm N}$ respectively.

The MOLspheric temperature-density distribution shows a slight decrease in temperature and $N_{\rm CO}$ from $T_{\rm mol}\big(\frac{R_{\rm mol,0}}{R_\star}\big)=960K$ to $T_{\rm mol}\big(\frac{R_{\rm mol,end}}{R_\star}\big)=958K$ and from $N_{\rm CO}\big(\frac{R_{\rm mol,0}}{R_\star}\big)=10^{21.5}mol/cm^{-2}$ to $N_{\rm CO}\big(\frac{R_{\rm mol,end}}{R_\star}\big)=10^{21.1}mol/cm^{-2}$ respectively.
We adjusted $\diameter_\star=24.7\pm 0.4mas$ to obtain the best fit between the model and the visibilities in the continuum, and according to \citet[][and references therein]{2014A&A...561A..47O} we fixed $\vmicromol=\vmicro=2\kms$ in order to obtain the best agreement between the width of modeled molecular lines and the observations.
The bottom portion of Fig.\,\ref{gamCru-Res} shows a comparison between the modeled and observed normalized flux $F/F_{\rm Continuum}(\lambda)$, visibilities $V(\lambda)$ and closure phases.
					
The modeled visibilities are of approximately the same amplitudes with respect to the observations at the continuum (Fig.~\ref{gamCru-Res}), only by adjusting the angular size ($\diameter_\star=24.7\pm 0.4mas$), which shows that the target is not perfectly spherical.

$\gamma$ Cru, which is an RGB star (as $\alpha$ Boo), and hence should host a MOLsphere/COMosphere \citep[of $T_{\rm mol}\sim 1000K$][]{2003ApJ...598..610A}. Instead, it presents a solution of a thin CO layer, located far from the photosphere, due to being pushed by an important stellar Alfv\'{e}n-wave-driven wind \cite{2007ApJ...659.1592S,2010ApJ...723.1210A}.

\subsection{$\lambda$ Vel}
\label{lamvel}
For $\lambda$ Vel, the best \textsc{PAMPERO} fit model shows a total thickness of $0.5R_\star$ for five layers each with a thickness of $0.1R_\star$. The temperature-density distribution is $T_0=1000K$ with a coefficient of $\zeta_{\rm T}=0.1$, and $N_0=10^{21.5}mol/cm^{-2}$ with a coefficient of $\zeta_{\rm N}=50$.
The first two plots in Fig.\,\ref{lamVel-Res} depict the temperature-density distribution $T_{\rm mol}(r/R_\star)$ and $N_{\rm CO}(r/R_\star)$, where both distributions start from $R_{\rm mol,0}=4.5R_\star$ to decrease to $R_{\rm mol,end}=5.0R_\star$ with the coefficients of $\zeta_{\rm T}$ and $\zeta_{\rm N}$ respectively.

The MOLspheric temperature-density distribution shows a decrease in temperature and $N_{\rm CO}$ from $T_{\rm mol}\big(\frac{R_{\rm mol,0}}{R_\star}\big)=1000K$ to $T_{\rm mol}\big(\frac{R_{\rm mol,end}}{R_\star}\big)=990K$ and from $N_{\rm CO}\big(\frac{R_{\rm mol,0}}{R_\star}\big)=10^{21.5}mol/cm^{-2}$ to $N_{\rm CO}\big(\frac{R_{\rm mol,end}}{R_\star}\big)=10^{19.2}mol/cm^{-2}$ respectively.
We adjusted $\diameter_\star=11.3mas$ to obtain the best fit between the model and the observed visibilities in the continuum, and we fixed $\vmicromol=\vmicro=2\kms$ in order to get the best agreement between the width of modeled molecular lines and the observations.
The bottom portion of Fig.\,\ref{lamVel-Res} shows a comparison between the modeled and observed normalized flux $F/F_{\rm Continuum}(\lambda)$, visibilities $V(\lambda)$ and closure phases.
					
The modeled visibilities are approximately the same in amplitudes with respect to the observations at the continuum (little bit less at the biggest baseline $B=27.76m$, Fig.~\ref{lamVel-Res}), which means that the symmetry of our target is not perfectly spherical but the better than for $\gamma$ Cru.
We interpret the bad fit on the CO band head on the visibilities as a presence of a hot or a dark spot(s) on the MOLsphere and for which we cannot determine either the position or the number with our sparse $(u,v)$ coverage \citep[as demonstrated by][]{Miguel2014}.

$\lambda$ Vel, should be more an RGB than AGB star \citep{2003MNRAS.343L..79K}, just as $\alpha$ Boo and $\gamma$ Cru , and we expect a MOLsphere/COMosphere \citep[of $T_{\rm mol}\sim 1000K$][]{2003ApJ...598..610A}. Instead, it presents a solution with thick CO layer, which is far from the photosphere, because of an important stellar Alfv\'{e}n-wave-driven wind \cite{2007ApJ...659.1592S,2010ApJ...723.1210A}.

\subsection{$\alpha$ Sco}
\label{alpsco}
Our best \textsc{PAMPERO} fit model for $\alpha$ Sco shows a total thickness of $0.7R_\star$ for 7 layers each with a thickness of $0.1R_\star$. The temperature-density distribution is of $T_0=2350K$ with a coefficient of $\zeta_{\rm T}=1$, and $N_0=10^{21.5}mol/cm^{-2}$ with a coefficient of $\zeta_{\rm N}=35$.
The first two plots in Fig.\,\ref{alpSco-Res} depict the temperature-density distribution $T_{\rm mol}(r/R_\star)$ and $N_{\rm CO}(r/R_\star)$, where both distributions start from $R_{\rm mol,0}=1.06R_\star$ to decrease to $R_{\rm mol,end}=1.76R_\star$ with the coefficients of $\zeta_{\rm T}$ and $\zeta_{\rm N}$ respectively.
The MOLspheric temperature-density distribution varies from $T_{\rm mol}\big(\frac{R_{\rm mol,0}}{R_\star}\big)=2350K$ to $T_{\rm mol}\big(\frac{R_{\rm mol,end}}{R_\star}\big)=1900K$ with a stochastic distribution along $r/R_\star$ (with two peaks; $T_{\rm mol}(1.36R_\star)=3600K,\&\,T_{\rm mol}(1.66R_\star)=2200K$ and two floods; $T_{\rm mol}(1.16R_\star)=1850K,\&\,T_{\rm mol}(1.56R_\star)=2000K$, alternatively) and from $N_{\rm CO}\big(\frac{R_{\rm mol,0}}{R_\star}\big)=10^{21.5}mol/cm^{-2}$ to $N_{\rm CO}\big(\frac{R_{\rm mol,end}}{R_\star}\big)=10^{19.2}mol/cm^{-2}$ (with a flood of $N_{\rm CO}(1.46R_\star)=10^{16.63}mol/cm^{-2}$) respectively. Without this ``ad hoc'' manipulation for a $T_{\rm mol}$ and $N_{\rm co}$ of few layers, the first modeled visibility is high with respect to the observation.

We adjusted $\diameter_\star=37mas$ to obtain the best fit between the model and the observed visibilities in the continuum, and according to \cite[][and references therein]{2014A&A...561A..47O} we fixed $\vmicromol=\vmicro=5\kms$ in order to get the best agreement between the width of modeled molecular lines and the observations.
The bottom portion of Fig.\,\ref{alpSco-Res} shows a comparison between the modeled and observed normalized flux $F/F_{\rm Continuum}(\lambda)$, visibilities $V(\lambda)$ and closure phases.

The modeled visibilities are of approximately the same amplitudes with respect to the observations at the continuum (Fig.~\ref{alpSco-Res}), which means that the target is quasi-spherical as demonstrated on the recent velocity-resolved images of Antares \citep{2017Natur.548..310O}.

$\alpha$ Sco, as an RSG, its photosphere hosts large granules \citep[instead of 2-10 million present on the sun, as predicted by][]{1975ApJ...195..137S}. These granules were suspected to be caused principally by convection phenomenon \citep{2009A&A...508..923H,2009A&A...503..183O,2010A&A...515A..12C,2011A&A...528A.120C,2011A&A...529A.163O,2014A&A...572A..17M}, but \citet{2017Natur.548..310O} recently deduced that this phenomenon of upwelling and down-drafting motions cannot explain alone the atmospheric extension and turbulent motions observed on Antares. Indeed, \citet{2017Natur.548..310O} deduced by its images a MOLsphere size of $R_{\rm mol}\sim1.7R_\star$ with velocities of $V_{\rm mol}\sim -10\,\,to\,+20\kms$, indicating that the observed atmospheric extension and its density are much larger than theoretically predicted by \cite{2015A&A...575A..50A}.
Our Antares results (Fig.~\ref{alpSco-Res} \& Tab.~\ref{Tab4}) confirm the recent MOLspheric CO size $R_{\rm mol}\sim1.7R_\star$ of \citet{2017Natur.548..310O} and its stochastic behavior, with upwelling and down-drafting motions, as we can deduce from our temperature-density distributions. Therefore, we suggest a convective MOLsphere, just after the stellar convective area and its granules (by continuous manner along the stellar radii) to interpret our results. Indeed, only a MOLsphere with a convective behavior could explain this kind of temperature-density distributions $T_{\rm mol}(r/R_\star)$ \& $N_{\rm CO}(r/R_\star)$. To confirm our findings further work is needed using theoretical models of RSG \citep[e.g., 3 Dimensional convective models of][]{2011A&A...528A.120C, 2015A&A...575A..50A}.
We have to keep in mind that our solution is a rough estimation of the MOLsphere of $\alpha$ Sco, because it's very difficult to model the stochasity of the temperature-density distribution. What we present here is only the mean values of $T_{\rm mol}(r/R_\star)$ \& $N_{\rm CO}(r/R_\star)$. In addition, we interpret the poorer fit on the CO band head of the second and third visibility due to the presence of a hot or a dark spot(s) on the MOLsphere. We are unable to determine the position and number of spots with our sparse $(u,v)$ coverage \citep[as demonstrated by][]{Miguel2014}.

\subsection{W Hya}
\label{whya}
W Hya is a Mira star with magnitude variability $\Delta V=7.7-11.6$ for a cyclic period of 361 days \citep[GCVS 5.1;][]{2017ARep...61...80S}. We study the HR-K band AMBER data of this star at two luminosity phases; post-minimum light (phase 0.59) and pre-maximum light (phase 0.77) respectively.

The best fit\textsc{PAMPERO} models for W Hya (phases 0.59 and 0.77) found a total thickness of $0.4R_\star$ containing four layers each with a $0.1R_\star$thickness. We found temperature-density distributions of $T_0=1950K$ with a coefficient of $\zeta_{\rm T}=0.6$, and $N_0=10^{22.7}mol/cm^{-2}$ with a coefficient of $\zeta_{\rm N}=10$ at post-minimum light, and $T_0=2000K$ with a coefficient of $\zeta_{\rm T}=0.6$, and $N_0=10^{22.7}mol/cm^{-2}$ with a coefficient of $\zeta_{\rm N}=10$ at the pre-maximum.
The first two plots in Figs.\,\ref{whya-Res} and \ref{whya2-Res} depict the temperature-density distribution $T_{\rm mol}(r/R_\star)$ and $N_{\rm CO}(r/R_\star)$, which starts from $R_{\rm mol,0}=1.3R_\star$ and decreases to $R_{\rm mol,end}=1.7R_\star$ for the post-minimum phase. While for the pre-maximum phase the temperature-density distribution $T_{\rm mol}(r/R_\star)$ and $N_{\rm CO}(r/R_\star)$ starts from $R_{\rm mol,0}=1.4R_\star$ and decreases to $R_{\rm mol,end}=1.8R_\star$, both with the coefficients of $\zeta_{\rm T}$ and $\zeta_{\rm N}$ respectively.

The MOLspheric thermal distribution of the phase 0.59 decreases from $T_{\rm mol}\big(\frac{R_{\rm mol,0}}{R_\star}\big)=1950K$ to $T_{\rm mol}\big(\frac{R_{\rm mol,end}}{R_\star}\big)=1660K$, while the phase 0.77 decreases from $T_{\rm mol}\big(\frac{R_{\rm mol,0}}{R_\star}\big)=2000K$ to $T_{\rm mol}\big(\frac{R_{\rm mol,end}}{R_\star}\big)=1720K$, when the MOLspheric CO density distributions of the both luminosity phases decrease from $N_{\rm CO}\big(\frac{R_{\rm mol,0}}{R_\star}\big)=10^{22.7}mol/cm^{-2}$ to $N_{\rm CO}\big(\frac{R_{\rm mol,end}}{R_\star}\big)=10^{21.6}mol/cm^{-2}$.

Our best fit models for these two luminosity phases, show a clumpy MOLsphere with a total thickness $\sim10\%$ greater at post-minimum light ($\diameter_\star=43.75\pm0.75mas$, $R_{\rm mol,end}=1.7R_\star$) than at the pre-maximum one ($\diameter_\star=39mas$, $R_{\rm mol,end}=1.8R_\star$). The MOLspheric temperature of the first layer is slightly higher at the pre-maximum ($T_0=2000K$) than at the post-minimum ($T_0=1950K$) -almost the same $T_0$ taking into account the uncertainty of $\pm50K$ for both-, while the CO column density and MOLspheric temperature coefficients appear to stay the same ($N_0=10^{22.7}mol/cm^{-2}$, $\zeta_{\rm N}=10$ \& $\zeta_{\rm T}=0.6$). In another term, the thermal distribution seems to be higher at the phase 0.77 than at the phase 0.59, while the CO density distribution appears to stay the same.
The width of modeled CO lines agrees with the observations, with a micro-turbulent velocity of $\vmicromol=3.7\kms$ for the phase 0.59, while for the phase 0.77, the value of $4\kms$ seems to be the optimum $\vmicromol$, which means that the star's activity also affects the micro-turbulent velocity, implying that the activity and micro-turbulent velocity are proportional to each other.

We perturbed $\diameter_\star=43.75mas$ by $\pm0.75mas$ at the post-minimum, to obtain a good fit between the amplitudes of the three modeled visibilities compared with the observed one, where we fixed $\diameter_\star=39mas$ at the pre-maximum, which implies that W Hya seems to be more spherical at the phase 0.59 than at phase 0.77. We interpret this as the light emitted by W Hya at the maximal activity disrupt the symmetry of the MOLsphere material obtained during the minimal activity due to the gravitational force.
We observe among on all our sample of evolved stars, that W Hya has the highest column density with $N_{\rm CO}=10^{22.7}mol/cm^{-2}$ and the lowest coefficient $\zeta_{\rm N}=10$. We explain this behavior by the fact that W Hya is a Mira variable star with a regular cyclic activity which feeds continuously (every 361 days) the interstellar medium.

We interpret the poor fit on the CO band head (and some other individual lines) with the same visibilities (on the both phases) due to the presence of a hot or a dark spot(s) on the MOLsphere.  We cannot determine the position or the number of spots with our sparse $(u,v)$ coverage \citep[as demonstrated by][]{Miguel2014}.

Our average MOLspheric temperature and CO MOLspheric sizes of W Hya agree well with the results of a prior study which found $T_{\rm mol}\sim 1500K$, $R_{\rm mol}=1.9-3.0R_\star$ \citep{2016A&A...589A..91O}.

\section{Conclusions}
\label{conclu}
Using the Differential Interferometry technique of VLTI/AMBER, we were able to spatially resolve, the individual CO first overtone lines, of a sample of different evolved stars, namely; the Red SuperGiant $\alpha$ Scorpii, the Red Giant Branch stars $\alpha$ Bootis and $\gamma$ Crucis, the K giant $\lambda$ Velorum, the normal M-giants BK Virginis and SW Virginis, and the Mira variable star W Hydrae at two different luminosity phases.  
The uniform-disk diameters in the CO lines are distinctly higher compared to the continuum.
Despite this, the MARCS photospheric model reproduces, in an acceptable manner, the spectra of our sample of evolved stars in the observed CO lines. However, the predictions of modeled angular size in CO lines and particularly in the band head remain largely underestimated. This reveals more extended CO layers than predicted by the MARCS photosphere model alone.
Our CO-multilayer model, combined with the MARCS model, satisfactorily explains the specto-interferometric observations, namely: the spectra, visibilities and $\Psi$ for several kinds of evolved stars.
The deduced CO temperature-density distributions are equal or greater than, the uppermost layer of the photospheres. Therefore, some heating mechanisms should exist in the outer atmospheres of the evolved stars, which requires further investigation.

While every star is unique, our results reveal that for the different kinds of evolved stars present, in general, the same temperature-density distributions for their spectral type. Indeed, our \textsc{PAMPERO} model, with its continuous and multilayer MOLsphere approach, shows that:
\begin{itemize}
\item The RSG Antares presents a convective MOLsphere directly after its convective photosphere, from $1.06R_\star$ to $1.76R_\star$, with a highest $T_{\rm mol}=2350-1900K$ combined with a lower $N_{\rm CO}=10^{21.5}-10^{19.2}mol/cm^{-2}$.
\item The RGB stars $\alpha$ Bootis, $\gamma$ Crucis and the K giant $\lambda$ Velorum \citep[confirming the results of][that $\lambda$ Velorum is more RGB than AGB]{2003MNRAS.343L..79K}, present an important gap between the outer photosphere and the inner MOLsphere of $2.5\,to\,5.0R_\star$ with a size of $0.1\,to\,0.5R_\star$ for $T_{\rm mol}\sim 2000-1000K$ and $N_{\rm CO}\sim 10^{21.5}-10^{17}mol/cm^{-2}$. 
\item The normal M-giants BK Vir and SW Vir, which may have a small gap between the outer photosphere and the inner MOLsphere of $\sim0.05R_\star$, have a MOLsphere with a large size of $2\,to\,3R_\star$ for $T_{\rm mol}\sim2000-1500K$ and $N_{\rm CO}\sim 10^{22.5}-10^{10}mol/cm^{-2}$. 
\item The Mira variable star, W Hya presents a moderate gap between the outer photosphere and the inner MOLsphere of $\sim0.2-0.3R_\star$ with a size of $\sim0.4R_\star$ for $T_{\rm mol}\sim2000-1500K$ and a rich $N_{\rm CO}\sim 10^{22.7}-10^{21.5}mol/cm^{-2}$.
\end{itemize}

\begin{figure}
\centering
\includegraphics[width=1.0\hsize,draft=false]{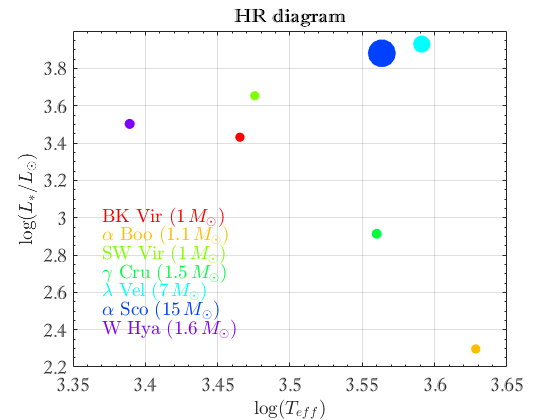}
\caption{HR diagram of all evolved stars studied in this work, where the symbol size of each star is proportional to its mass.}\label{HR-Sample_EvolvedStars2D}
\end{figure}

Fig.~\ref{HR-Sample_EvolvedStars2D} summarizes the HR diagram according to stellar mass of all our targets while the last part of Table\,\ref{Tab4} recaps the best six free parameters ($T_0,\,N_0,\,\zeta_{\rm T},\,\zeta_{\rm N},\,R_{\rm mol,0},\,\&\,R_{\rm mol,end}$) combined with their associated uncertainties from our modeling of the eight evolved stars.

We have demonstrated that our continuous and multilayer MOLsphere approach, can be used to study a large panel of evolved stars and that bi-layer MOLspheric models could be misleading, especially in the case of an observed continuous MOLsphere. Indeed, our multilayer model is a refinement of the previous bi-layer models. In this work we studied only the temperature-density of the CO molecule observed by VLTI/AMBER, but using other instruments such as CHARA/VEGA we will be able to study the Titanium monoxide (TiO) spectral lines, not only for regular evolved stars but also for Yellow Hyper-Giants (YHG). This would also allow the study of MOLspheric temperature-density for other molecules such as water vapor ($H_20$) and Silicon monoxide (SiO), simultaneously with the CO molecule. A MOLspheric temperature-density study of dust around evolved stars, especially in the L, M and N bands with the new instrument VLTI/MATISSE \citep{2014Msngr.157....5L} would be very interesting too.
For future work we would use the Markov Chain Monte Carlo (MCMC) method on the free parameters and their uncertainties for our \textsc{PAMPERO} model \citep[as we did in][]{2018MNRAS.480.1263H}. Prior to this, we first have to find a technical solution on how to efficiently use the MCMC method for large intervals of the six free parameters within reasonable calculation times. An eventual solution is to develop a global optimizer which is a hybrid between MCMC and a genetic algorithm, called GEMC \citep[Genetic Evolution Markov Chains, e.g.][]{2013MNRAS.428.3671T,2015MNRAS.450.1760T}

This work, in addition to reinforcing that AMBER observations with high-spectral resolution are efficient to constrain the physical properties of the outer atmosphere of cool evolved stars, strengthen the existence of some heating mechanism in the outer atmosphere of AGBs, RGBs and RSGs.
This work shows that evolved stars deserve further study using IR spectro-interferometry.
We advocate for a large survey, at different wavelength and using a richer $(u,v)$-coverage, for different ranges of temperatures and luminosities, of these stars. This would better help us, understand the mass-loss mechanism of the big family of evolved stars.


\bibliographystyle{mnras}
\bibliography{Biblio_pampero}


\appendix

\section{Data interpretation figures}
\label{Data_int_Fig}

We gathered here all the figures of the data interpretation section (Sec.~\ref{Data_int}), namely: $\gamma$ Cru, $\lambda$ Vel, $\alpha$ Sco and W Hya (pahse 0.59 and 0.77) respectively.

\begin{figure*}
\centering
\includegraphics[width=0.44\hsize,draft=false]{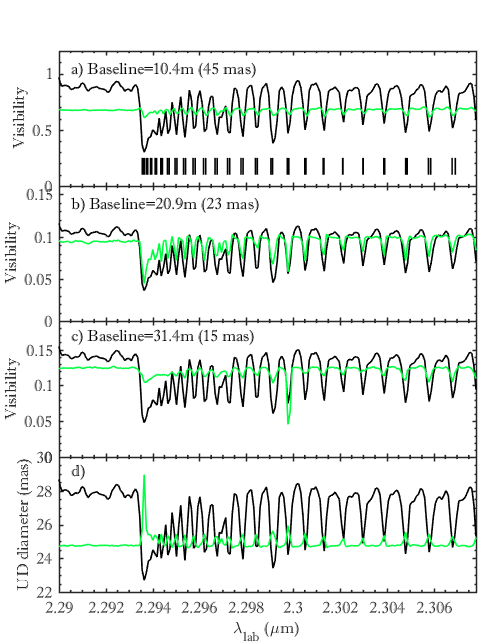}
\includegraphics[width=0.44\hsize,draft=false]{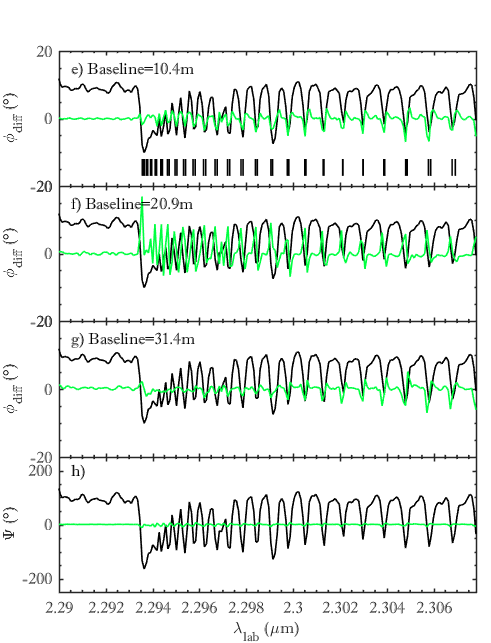}
\caption{AMBER data of $\gamma$ Cru. In each panel, the scaled observed spectrum is plotted by the black solid lines. a)-c) Visibilities observed on the reported triplet baselines are shown on a), b) and c) plots (colored lines). The corresponding spatial resolutions are also given. d) Uniform-disk diameter (colored line) derived by fitting the visibilities shown in panels a)-c). e)-g) Differential phases ($\phidiff$) observed on the reported triplet baselines are shown on e), f) and g) plots (colored lines). h) Closure phase ($\Psi$) in colored line also.} \label{gamcru-Data}
\end{figure*}

\begin{figure*}
\centering
\includegraphics[width=0.44\hsize,draft=false]{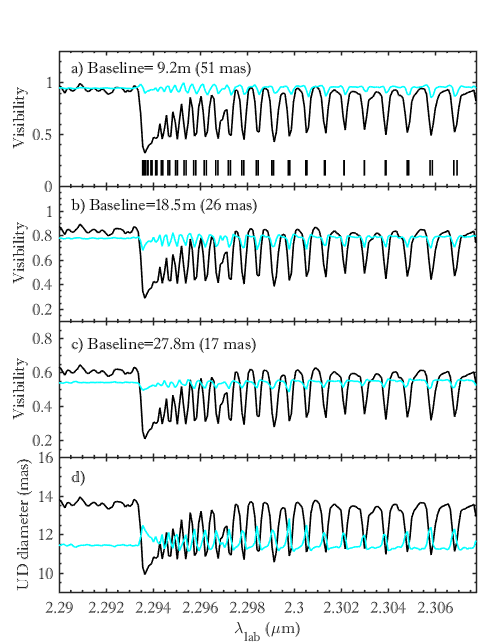}
\includegraphics[width=0.44\hsize,draft=false]{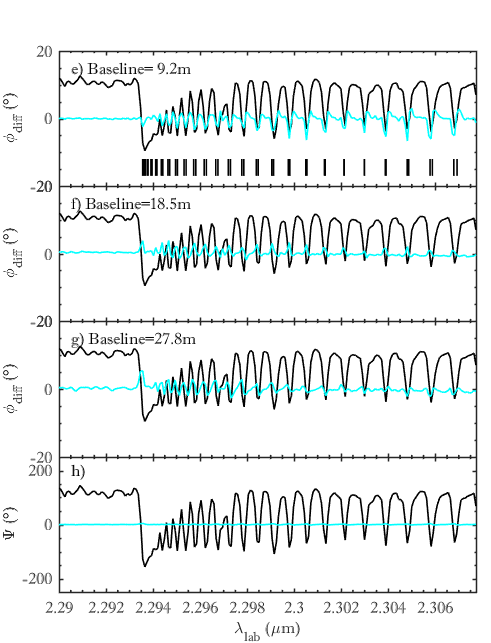}
\caption{Same as Fig.~\ref{gamcru-Data} for $\lambda$ Vel.} \label{lamvel-Data}
\end{figure*}

\begin{figure*}
\centering
\includegraphics[width=0.44\hsize,draft=false]{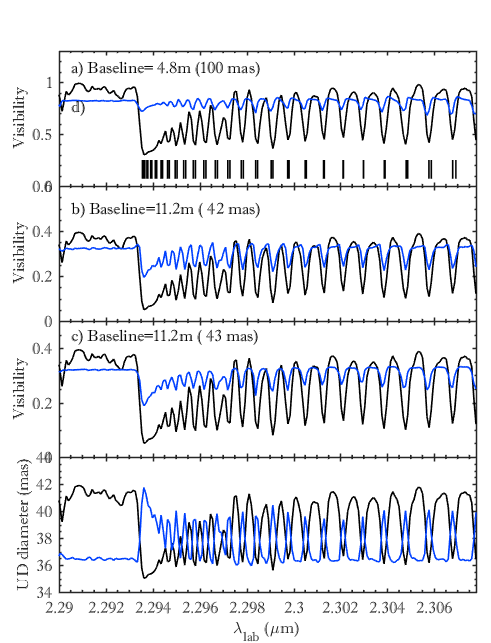}
\includegraphics[width=0.44\hsize,draft=false]{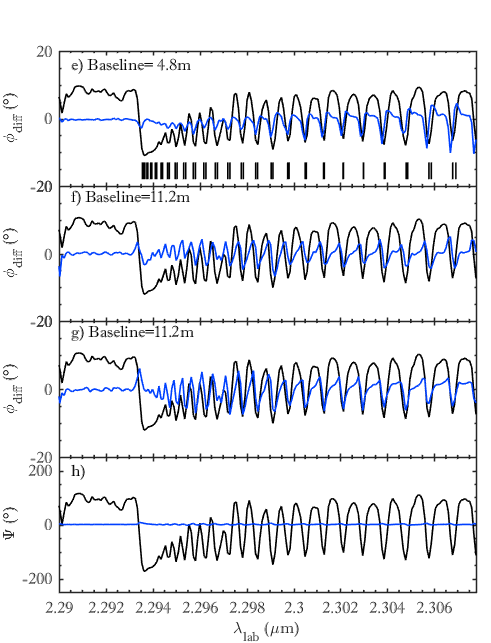}
\caption{Same as Fig.~\ref{gamcru-Data} for $\alpha$ Sco.} \label{alpsco-Data}
\end{figure*}

\begin{figure*}
\centering
\includegraphics[width=0.44\hsize,draft=false]{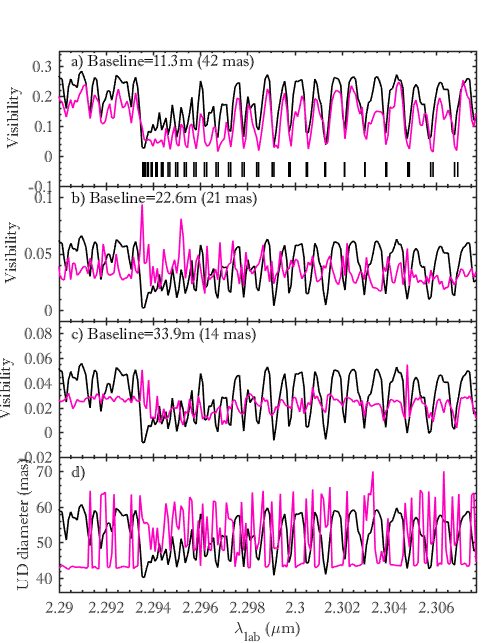}
\includegraphics[width=0.44\hsize,draft=false]{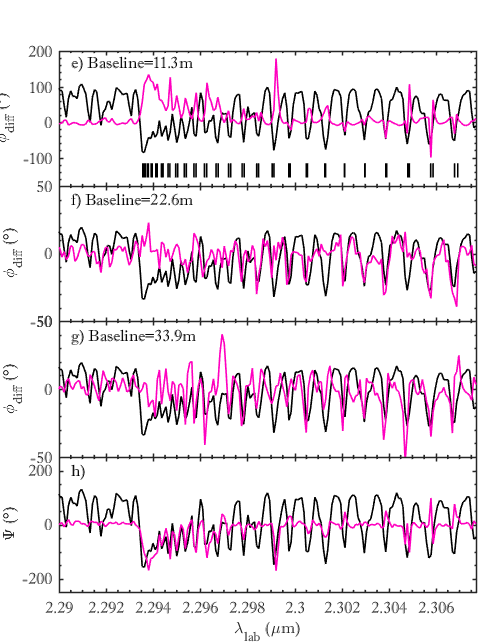}
\caption{Same as Fig.~\ref{gamcru-Data} for of W Hya (phase 0.59).} \label{whya-Data}
\end{figure*}

\begin{figure*}
\centering
\includegraphics[width=0.44\hsize,draft=false]{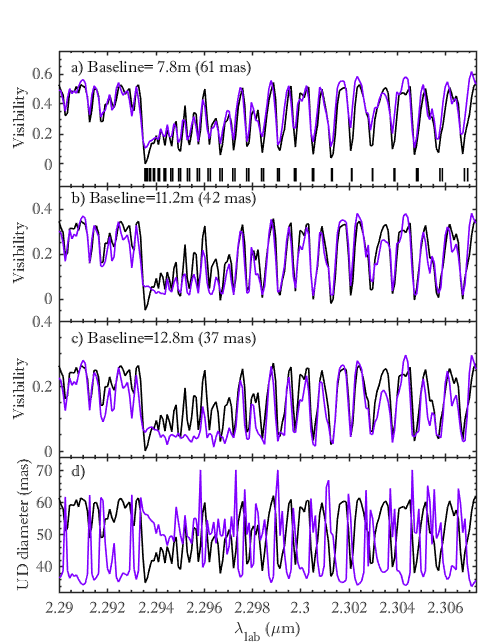}
\includegraphics[width=0.44\hsize,draft=false]{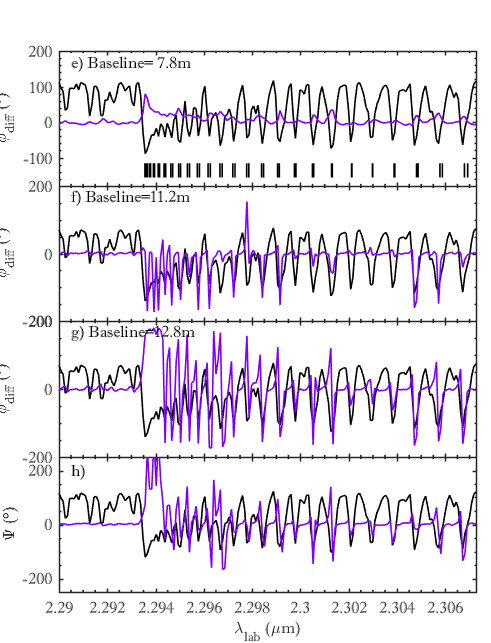}
\caption{Same as Fig.~\ref{gamcru-Data} for of W Hya (phase 0.77).} \label{whya2-Data}
\end{figure*}

\section{Parameters of stellar atmosphere models and MOLsphere's results}
\label{Res_Tab}

\begin{figure}
\centering
\includegraphics[width=1.0\hsize,draft=false]{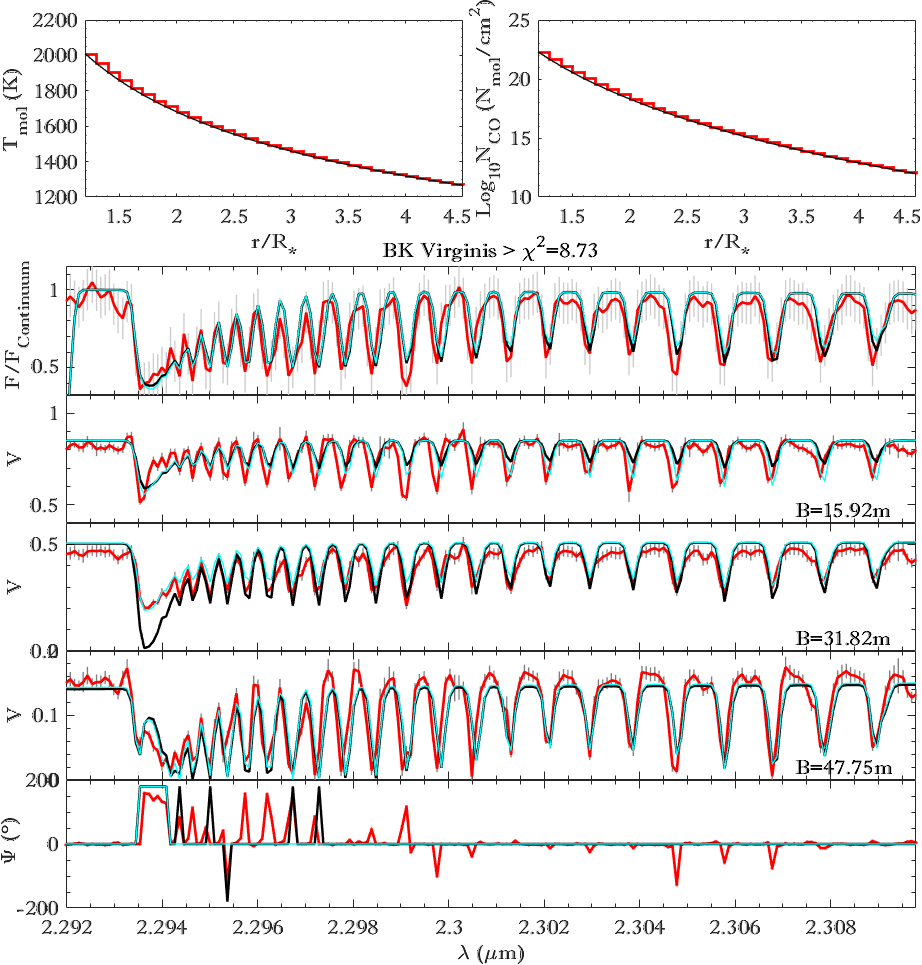}
\caption{Our best \textsc{PAMPERO} model for BK Vir. \textbf{Top:} MOLsphere's temperature ($T_{\rm mol}$) and CO column density ($N_{\rm CO}$) distributions along the stellar radius $(r/R_\star)$ respectively (layer by layer in colour stair shape lines and the global behavior in black lines). \textbf{Bottom:} Spectro-interferometric comparison between the observation (the line whose colour is used in the top panel) and our best model (black line), namely the normalized flux $F/F_{\rm Continuum}(\lambda)$, the visibilities $V(\lambda)$ for their respective baselines ($B$) and closure phase $\Psi(\lambda)$. In another colour we show the best bi-layers result (of Tab.~\ref{Tab4}).}\label{BKVir-Res}
\end{figure}

\begin{figure}
\centering
\includegraphics[width=1.0\hsize,draft=false]{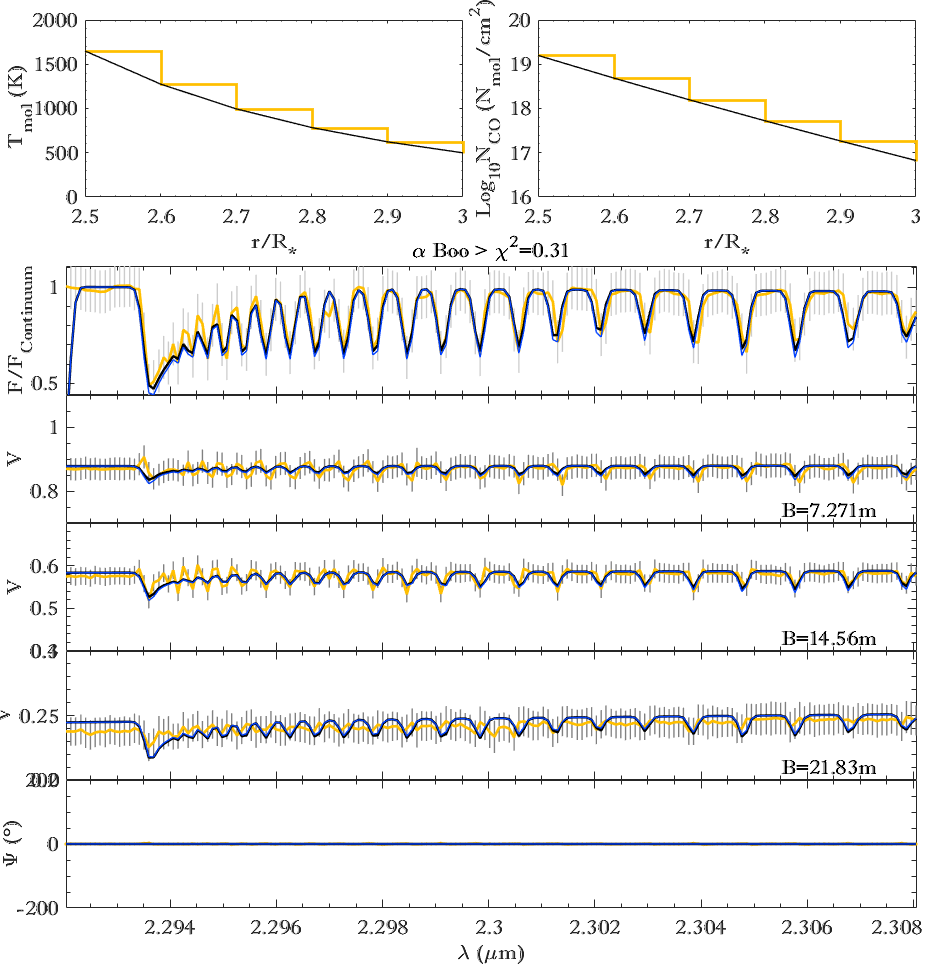}
\caption{Our best \textsc{PAMPERO} model for $\alpha$ Boo in the same manner as the Fig.~\ref{BKVir-Res}.}\label{alpBoo-Res}
\end{figure}

\begin{figure}
\centering
\includegraphics[width=1.0\hsize,draft=false]{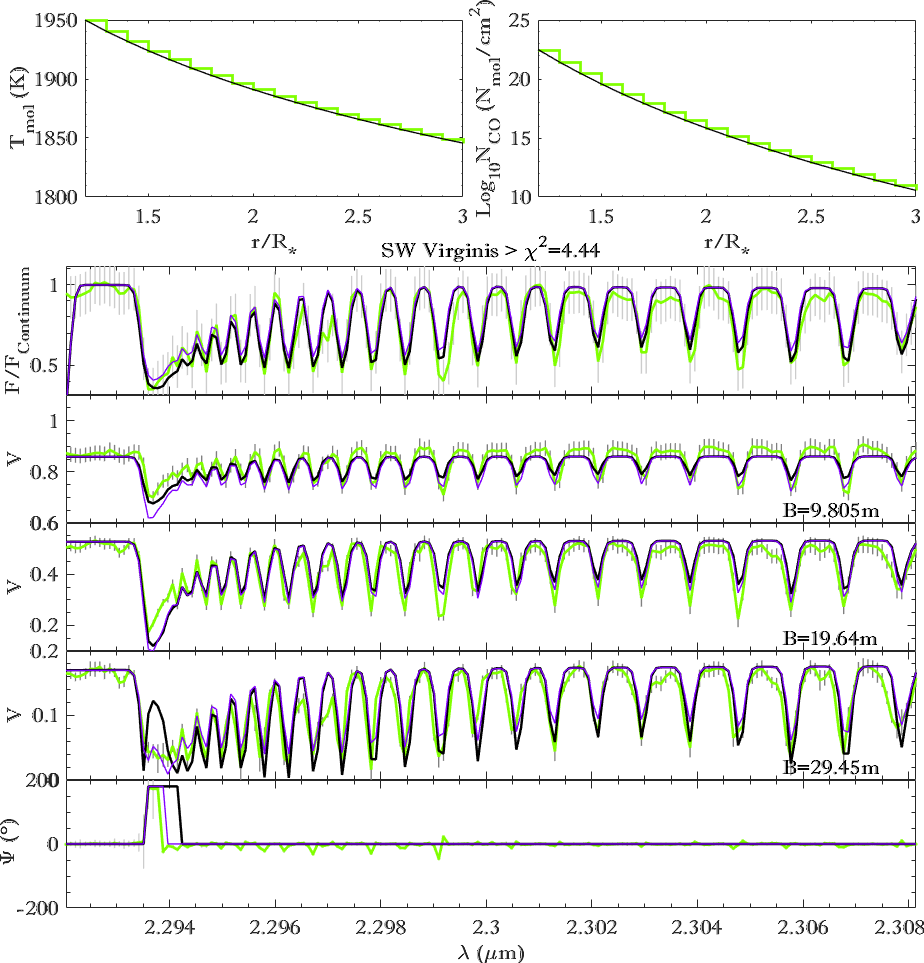}
\caption{Our best \textsc{PAMPERO} model for SW Vir in the same manner as the Fig.~\ref{BKVir-Res}.}\label{SWVir-Res}
\end{figure}

We summarize, on a big table, the basic stellar parameters, the chosen MARCS models and the best six free parameters with their uncertainties of \textsc{PAMPERO} for our sample of eight evolved stars, where we compare old published results of BK Vir, $\alpha$ Boo and SW Vir (using a MARCS+2-layer  model) with ours (\textsc{PAMPERO}). Note that \textsc{PAMPERO} works also with two layers and it gives the same results that are found in the literature, with negligible differences, as shown on Figs.~\ref{BKVir-Res} to \ref{SWVir-Res}.

\begin{table*}
\caption{Basic stellar parameters, chosen MARCS models and the best six free parameters with their uncertainties of \textsc{PAMPERO} for our sample of evolved stars.} \label{Tab4}
\begin{sideways}
\centering
\begin{threeparttable}
\centering 
\begin{tabular}{ccccccccc}
\hline \hline
\textbf{Star} & \textbf{BK Vir} & \textbf{$\alpha$ Boo} & \textbf{SW Vir} & \textbf{$\gamma$ Cru} & \textbf{$\lambda$ Vel} & \textbf{$\alpha$ Sco} & \multicolumn{2}{|c|}{\textbf{W Hya}} \\
\hline
\textbf{Spectral type} & M7III & K1.5III & M7III & M3.5III & K4Ib & M1.5Ib & \multicolumn{2}{|c|}{M7.5e}\\
              & (AGB) & (RGB) & (AGB) & (RGB) & (RGB/AGB) & (RSG) & \multicolumn{2}{|c|}{(Mira)} \\
\textbf{Variability type}$^{(\dagger)}$ & SRB & No & SRB & No & LC & LC & \multicolumn{2}{|c|}{SRA} \\
\textbf{Magnitude$^{(\dagger)}$ (V)} & $7.28-8.8$ & $-0.05^{(a)}$ & $6.40-7.90$ & $1.64^{(a)}$ & $2.14-2.30$ & $0.88-1.16$ & \multicolumn{2}{|c|}{$7.7-11.6$}\\
\textbf{distance$^{(\ddagger)}$ (pc)} & $181^{+25}_{-20}$ & $11.26\pm0.07$ & $143^{+19}_{-15}$ & $27.15\pm0.13$ & $167\pm3$ & $169.78^{+34.72}_{-24.64}$ & \multicolumn{2}{|c|}{$78^{+6.5}_{-5.6}$$^{(b)}$}\\
$\boldsymbol \diameter_\star$ (mas) & $10.73\pm0.23$ & $20.4\pm0.2$ & $16.23\pm0.20$ & $24.70\pm0.35$ & $11.1\pm0.8$ & $37.61\pm0.12$ & \multicolumn{2}{|c|}{$46.6\pm0.1$}\\
$\boldsymbol \Tmean$ \textbf{(K)} & $2920\pm150$ & $4250\pm50$ & $2990\pm50$ & $3630\pm90$ & $3800-4000$ & $3660\pm120$ & \multicolumn{2}{|c|}{2400-2500}\\
$\boldsymbol \log g$ $(cm.s^{-2})$ & $\sim-0.17$ & $+1.7\pm0.1$ & $-0.3\pm0.1$ & $+0.9\pm0.1$ & $+0.64$ & $-0.2\pm0.3$ & \multicolumn{2}{|c|}{$-0.86\pm0.17$$^{(\ast)}$}\\
$\boldsymbol M_\star$ $\boldsymbol (\Msun)$ & $\sim 1$ & $\sim 1.1$ & $1-1.25$ & $1.5\pm0.3$ & $7\pm1$ & $15\pm5$ & \multicolumn{2}{|c|}{$1.6\pm0.4$$^{(\ast)}$}\\
$\boldsymbol L_\star$ $\boldsymbol (\Lsun)$ & $2700\pm850$ & $198\pm3$ & $4500\pm1100$ & $820\pm80$ & $8511\pm982$ & $7600^{+5300}_{-3100}$ & \multicolumn{2}{|c|}{$3180^{+550}_{-440}$}\\
$\boldsymbol \vmicro$ $\boldsymbol (\kms)$ & $3-4$ & $\sim 2$ & $\sim 4$ & $\sim 2$ & $\sim 2$ & $\sim 5$ & \multicolumn{2}{|c|}{$3-4$}\\
                         & (no solar) & (solar) & (no solar) & (solar) & (solar) & (no solar) & \multicolumn{2}{|c|}{(no solar)}\\
$\boldsymbol \FeH$ & $\sim 0.0$ & $-0.5$ & $\sim 0.0$ & $\sim 0.0$ & $0.06$ & $\sim 0.0$ & \multicolumn{2}{|c|}{$0.78$$^{(\ast)}$}\\
                   & (solar) & (no solar) & (solar) & (solar) & (solar) & (solar) & \multicolumn{2}{|c|}{(no solar)}\\
\textbf{Chemical composition} & $C/N$ & $\ccrate$ & $\corate$ & $\corate$ & $C/O$ & $N/O$ & \multicolumn{2}{|c|}{$\corate$}\\
\textbf{} & $1.5$ & $7\pm2^{(c)}$ & $18^{(d)}$ & $\sim 20$ & $\sim 0.9^{(e)}$ & $\sim 1^{(f)}$ & \multicolumn{2}{|c|}{$10^{(d)}$}\\
\textbf{References} & (1) & (2) \& (3) & (4) & (3) & (5) & (3),(6) \& (7) & \multicolumn{2}{|c|}{(8) \& (9)}\\
\hline
\multicolumn{8}{|c|}{\textbf{MARCS model}}\\
$\boldsymbol \Tmean$ & \textbf{3000} & \textbf{4250} & \textbf{3000} & \textbf{3600} & \textbf{4000} & \textbf{3600} & \multicolumn{2}{|c|}{\textbf{2500}}\\
$\boldsymbol \log g$ & \textbf{0.0} & \textbf{1.5} & \textbf{0.0} & \textbf{1.0} & \textbf{1.0} & \textbf{0.0} & \multicolumn{2}{|c|}{\textbf{0.0}}\\
$\boldsymbol M_\star$ & \textbf{1.0} & \textbf{1.0} & \textbf{1.0} & \textbf{1.0} & \textbf{5.0} & \textbf{5.0} & \multicolumn{2}{|c|}{\textbf{1.0}}\\
$\boldsymbol \vmicro$ & \textbf{2.0} & \textbf{2.0} & \textbf{2.0} & \textbf{2.0} & \textbf{2.0} & \textbf{5.0} & \multicolumn{2}{|c|}{\textbf{2.0}}\\
$\boldsymbol \FeH$ & \textbf{+0.0} & \textbf{-0.5} & \textbf{+0.0} & \textbf{+0.0} & \textbf{+0.0} & \textbf{+0.0} & \multicolumn{2}{|c|}{\textbf{+0.25}}\\
\textbf{CN-cycled composition} & \textbf{moderately} & \textbf{moderately} & \textbf{moderately} & \textbf{moderately} & \textbf{moderately} & \textbf{heavily} & \multicolumn{2}{|c|}{\textbf{heavily}}\\
\hline
\multicolumn{8}{|c|}{\textbf{In the literature $\boldsymbol >$ MARCS + 2-layer MOLsphere models -with $0.1R_\star$ of thickness- $ ^{\boldsymbol\rm (Reference)}$}}\\
\textbf{$R_{\rm inner}$ $(R_\star)$} & $1.2–1.25$ $^{(1)}$ & $1.04\pm0.02$ $^{(2)}$ & $1.3\pm0.1$ $^{(4)}$ & - & - & - & - & - \\
\textbf{$R_{\rm outer}$ $(R_\star)$} & $2.5–3.0$ $^{(1)}$ & $2.6\pm0.2$ $^{(2)}$  & $2.0\pm0.2$ $^{(4)}$ & - & - & - & - & - \\
\textbf{$T_{\rm inner}$ $(K)$} & $1900–2100$ $^{(1)}$ & $1600\pm400$ $^{(2)}$  & $2000\pm100$ $^{(4)}$ & - & - & - & - & - \\
\textbf{$T_{\rm outer}$ $(K)$} & $1500–2100$ $^{(1)}$ & $1800\pm100$ $^{(2)}$  & $1700\pm100$ $^{(4)}$ & - & - & - & - & - \\
\textbf{$N_{\rm CO,inner}$ $(mol/cm^{-2})$} & $(1-2)\times10^{22}$ $^{(1)}$ & $10^{20\pm0.3}$ $^{(2)}$  & $10^{22\pm0.3}$ $^{(4)}$ & - & - & - & - & - \\
\textbf{$N_{\rm CO,outer}$ $(mol/cm^{-2})$} & $10^{19}-10^{20}$ $^{(1)}$ & $10^{19\pm0.15}$ $^{(2)}$  & $10^{20\pm0.6}$ $^{(4)}$  & - & - & - & - & - \\
\hline
\multicolumn{8}{|c|}{\textbf{MARCS + \textsc{PAMPERO} models (Thickness = $0.1R_\star$)}}\\
\textbf{Luminosity Phase} & - & - & - & - & - & - & \textbf{0.59} & \textbf{0.77}\\
$\boldsymbol R_{\rm mol,0}\,(R_\star)$ & $1.2\pm0.2$ & $2.5\pm0.2$ & $1.2\pm0.1$ & $5.15\pm0.1$ & $4.5\pm0.1$ & $1.06^{+0.05}_{-0.0}$ & $1.3\pm0.05$ & $1.4\pm0.05$ \\
$\boldsymbol R_{\rm mol,end} \,(R_\star)$ & $4.5\pm0.2$ & $3.0\pm0.2$ & $3.0\pm0.2$ & $5.25\pm0.1$ & $5.0\pm0.1$ & $1.76\pm0.05$ & $1.7\pm0.05$ & $1.8\pm0.05$ \\
$\boldsymbol T_0$ (K) & $2010\pm50$ & $1650\pm30$ & $1950\pm50$ & $960\pm40$ & $1000\pm50$ & $2350\pm50$ & $1950\pm50$ & $2000\pm50$ \\
$\boldsymbol \zeta_{\rm T}$ & $0.35\pm0.2$ & $6.56\pm0.50$ & $0.06\pm0.05$ & $0.10\pm0.05$ & $0.10\pm0.05$ & $1\pm0.1$ & $0.60\pm0.05$ & $0.60\pm0.05$ \\
$\boldsymbol N_0$ $(mol/cm^{-2})$ & $10^{22.3}\pm10^{0.2}$ & $10^{19.2}\pm10^{0.1}$ & $10^{22.5}\pm10^{0.2}$ & $10^{21.5}\pm10^{0.2}$ & $10^{21.5}\pm10^{0.2}$ & $10^{21.5}\pm10^{0.2}$ & $10^{22.7}\pm10^{0.2}$ & $10^{22.7}\pm10^{0.2}$ \\
$\boldsymbol \zeta_{\rm N}$ & $18\pm3$ & $30\pm5$ & $30\pm5$ & $50\pm5$ & $50\pm5$ & $35\pm5$ & $10\pm3$ & $10\pm3$ \\
\textbf{Number of layers} & 33 & 5 & 18 & 1 & 5 & 7 & 4 & 4 \\
\hline \hline

\end{tabular}
\begin{tablenotes}
		\scriptsize
		$(1)$ \citet{2012A&A...537A..53O} and references therein, 
		$(2)$ \citet{2018A&A...620A..23O} and references therein, 
		$(3)$ \citet{2014A&A...561A..47O} and references therein,\\ 
		$(4)$ \citet{2019A&A...621A...6O} and references therein,
		$(5)$ \citet{1999ApJ...521..382C} and references therein,\\
		$(6)$ \citet{2013A&A...555A..24O} and references therein,
		$(7)$ \citet{2017Natur.548..310O} and references therein,\\
		$(8)$ \citet{2016A&A...589A..91O} and references therein,
		$(9)$ \citet{2017A&A...597A..20O} and references therein.\\
		$(a)$ \citet{2002yCat.2237....0D},
		$(b)$ \citet{2003A&A...403..993K},
		$(c)$ \citet{2003A&A...400..709D} with CNO-abundances of $7.96\pm0.20,\,7.61\pm0.25\,\&\,8.68\pm0.20\,dex$ respectively,
		$(d)$ \citet{2014A&A...566A.145R},
		$(e)$ \citet{2014AJ....147..137L},
		$(f)$ \citet{2006ApJ...645.1448T} with CNO-abundances of $\sim 8,\,\sim 8.5\,\&\,8.5\,dex$ respectively.\\
		$(\dagger)$ GCVS version 5.1 \citet{2017ARep...61...80S},
		$(\ddagger)$ \citet{2007A&A...474..653V}.\\
		$(\ast)$ Our estimation using the theoretical evolutionary tracks of \citet{2008A&A...484..815B}.\\
		\textbf{N.B.} Because of the similarity in $\Tmean,\,\log g$, and $M_\star$ between $\alpha$ Orionis and $\alpha$ Scorpii, according to \cite{2013A&A...553A...3O}, we adopt the same chemical composition for $\alpha$ Sco as that of $\alpha$ Ori.
\end{tablenotes}
\end{threeparttable}
\end{sideways}
\end{table*}

\section{MARCS + \textsc{PAMPERO} models grids and $\chi^2$ restricted minimization intervals}
\label{Grid_Tab}

We summarize, in the table below, the huge grids of MARCS + \textsc{PAMPERO} models and the $\chi^2$ restricted minimization intervals that we used to determine our best results (last box of Tab.~\ref{Tab4} above).

\begin{table*}
\caption{MARCS + \textsc{PAMPERO} models grids and $\chi^2$ restricted minimization intervals. The paths of our six free parameters are: $\Delta R=0.2R_\star$ for $R_{\rm mol,0}$ and $\Delta R=0.5R_\star$ for $R_{\rm mol,end}$, $\Delta T=100\,K$ for $T_0$, $\Delta N=10^{0.3}\,mol/cm^{-2}$ for $N_0$, $\Delta\zeta_{\rm T}=1$ and $\Delta\zeta_{\rm N}=10$.} \label{Tab6}
\centering
\begin{threeparttable}
\centering 
\begin{tabular}{cccccccc}
\hline \hline
\textbf{Star} & \textbf{BK Vir} & \textbf{$\alpha$ Boo} & \textbf{SW Vir} & \textbf{$\gamma$ Cru} & \textbf{$\lambda$ Vel} & \textbf{$\alpha$ Sco} & \textbf{W Hya} \\
\hline
\multicolumn{7}{|c|}{\textbf{MARCS + \textsc{PAMPERO} $>$ models grids}}\\
$\boldsymbol R_{\rm mol,0}\,(R_\star)$ & $1.2-1.6$ & $1.05-3.05$ & $1.2-1.6$ & $1.04-5.64$ & $1.02-5.02$ & $1.06-1.66$ & $1.15-1.55$ \\
$\boldsymbol R_{\rm mol,end} \,(R_\star)$ & $1.3-5.6$ & $1.1-4.1$ & $1.3-4.3$ & $1.2-6.2$ & $1.2-6.2$ & $1.2-3.2$ & $1.2-3.2$ \\
$\boldsymbol T_0$ (K) & $1900-2200$ & $1500-2500$ & $1900-2200$ & $900-2500$ & $900-2500$ & $1900-2500$ & $1800-2500$ \\
$\boldsymbol \zeta_{\rm T}$ & $0.01-5.01$ & $0.01-10.01$ & $0.01-5.01$ & $0.01-5.01$ & $0.01-5.01$ & $0.01-5.01$ & $0.01-5.01$ \\
$\boldsymbol N_0$ $(mol/cm^{-2})$ & $10^{21-23}$ & $10^{18-23}$ & $10^{21-23}$ & $10^{18-23}$ & $10^{18-23}$ & $10^{21-23}$ & $10^{21-23}$ \\
$\boldsymbol \zeta_{\rm N}$ & $5-25$ & $5-45$ & $5-45$ & $5-65$ & $5-65$ & $5-45$ & $5-25$ \\
\hline
\textbf{N$^{\circ}$ of grids} & $1$ & $3$ & $1$ & $3$ & $3$ & $1$ & $1$ \\
\textbf{N$^{\circ}$ of models} & $\sim 6500$ & $\sim 6500\times3$ & $\sim 8600$ & $\sim 12000\times3$ & $\sim 12000\times3$ & $\sim 21000$ & $\sim 17000$ \\
\textbf{Calculation time (hours)} & $\sim 60h$ & $\sim 180h$ & $\sim 72h$ & $\sim 300h$ & $\sim 300h$ & $180h$ & $\sim 150h$ \\
\hline
\multicolumn{7}{|c|}{\textbf{MARCS + \textsc{PAMPERO} models $>$ $\chi^2$ restricted minimization intervals}}\\
$\boldsymbol R_{\rm mol,0}\,(R_\star)$ & $1.2-1.3$ & $2-3$ & $1.2-1.3$ & $4.5-5.5$ & $4-5$ & $1.06-1.5$ & $1.25-1.45$ \\
$\boldsymbol R_{\rm mol,end} \,(R_\star)$ & $3-5$ & $2-5$ & $2-5$ & $5-6$ & $4.5-5.5$ & $1.5-2$ & $1.5-2$ \\
$\boldsymbol T_0$ (K) & $1900-2100$ & $1500-1800$ & $1900-2100$ & $900-1200$ & $900-1200$ & $2200-2500$ & $1900-2100$ \\
$\boldsymbol \zeta_{\rm T}$ & $0.1-1$ & $5-10$ & $0.01-0.1$ & $0.05-0.5$ & $0.05-0.5$ & $0.5-1.5$ & $0.1-1$ \\
$\boldsymbol N_0$ $(mol/cm^{-2})$ & $10^{22-23}$ & $10^{19-20}$ & $10^{22-23}$ & $10^{21-22}$ & $10^{21-22}$ & $10^{21-22}$ & $10^{22-23}$ \\
$\boldsymbol \zeta_{\rm N}$ & $15-25$ & $20-40$ & $20-40$ & $45-55$ & $45-55$ & $30-40$ & $5-15$ \\
\hline
\textbf{Calculation time (hours)} & $\sim 72h$ & $\sim 72h$ & $\sim 72h$ & $\sim 72h$ & $\sim 72h$ & $\sim 72h$ & $\sim 72h\times 2$ \\
\hline \hline
\end{tabular}
\end{threeparttable}
\end{table*}

\section{Results figures}
\label{Res_Fig}

We gathered here all the figures of the results section (Sec.~\ref{Res}), namely: BK Vir, $\alpha$ Boo, SW Vir, $\gamma$ Cru, $\lambda$ Vel, $\alpha$ Sco and W Hya (pahse 0.59 and 0.77) respectively.

\begin{figure}
\centering
\includegraphics[width=1.0\hsize,draft=false]{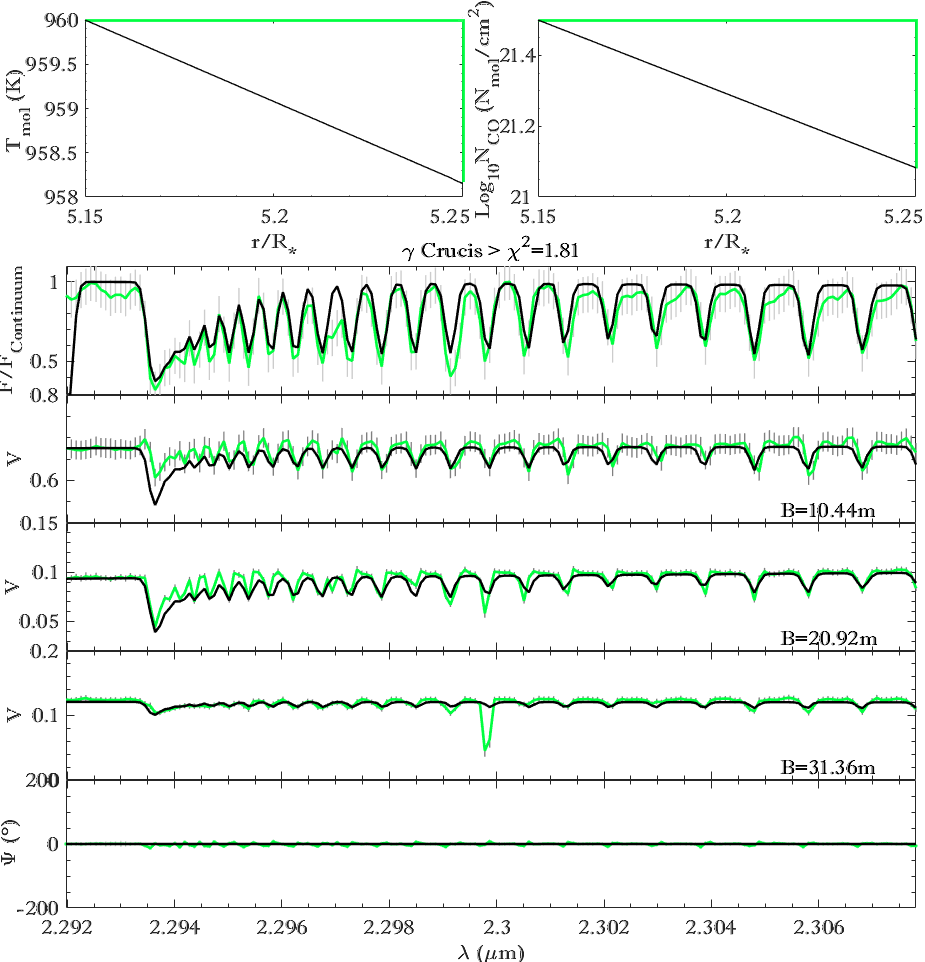}
\caption{Our best \textsc{PAMPERO} model for $\gamma$ Cru in the same manner as the Fig.~\ref{BKVir-Res}.}\label{gamCru-Res}
\end{figure}

\begin{figure}
\centering
\includegraphics[width=1.0\hsize,draft=false]{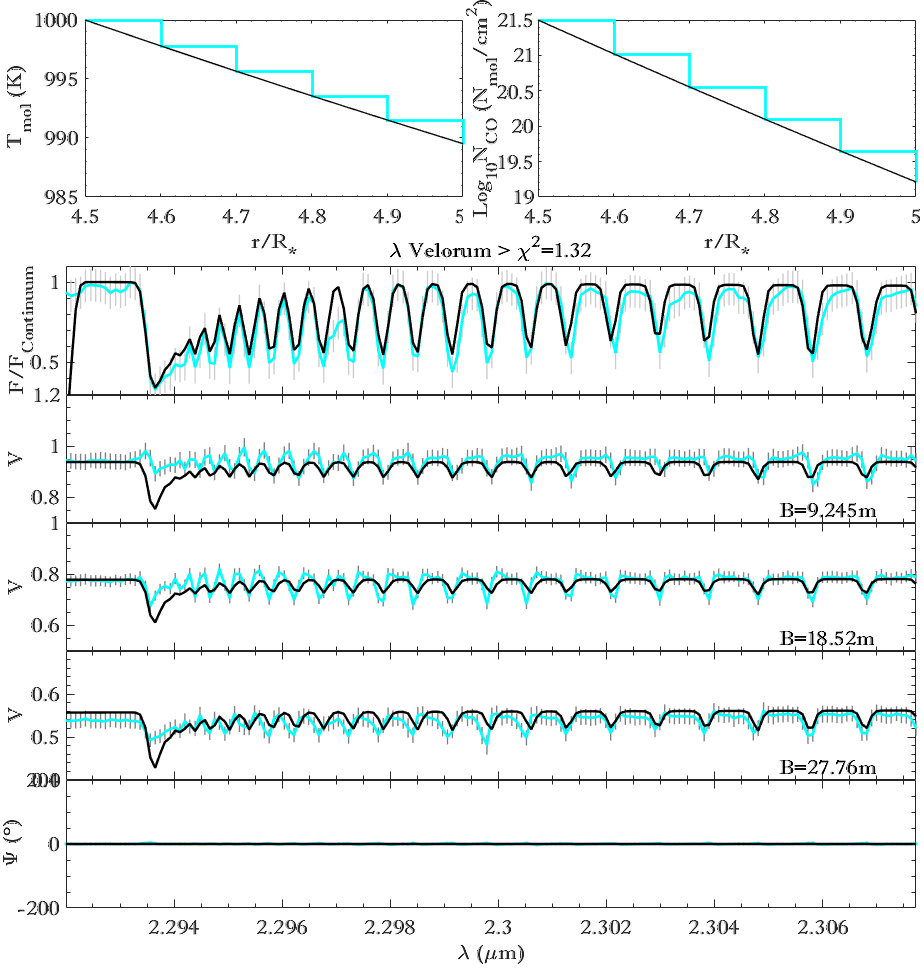}
\caption{Our best \textsc{PAMPERO} model for $\lambda$ Vel in the same manner as the Fig.~\ref{BKVir-Res}}\label{lamVel-Res}
\end{figure}

\begin{figure}
\centering
\includegraphics[width=1.0\hsize,draft=false]{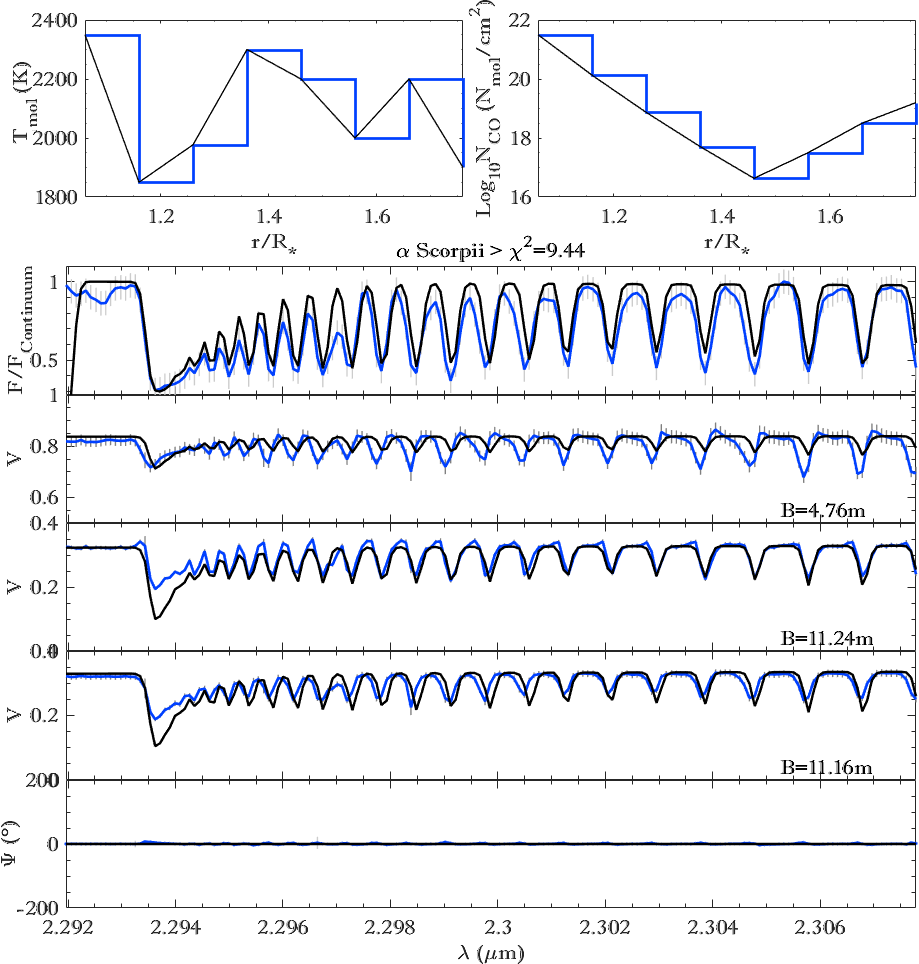}
\caption{Our best \textsc{PAMPERO} model for $\alpha$ Sco in the same manner as the Fig.~\ref{BKVir-Res}.}\label{alpSco-Res}
\end{figure}

\begin{figure}
\centering
\includegraphics[width=1.0\hsize,draft=false]{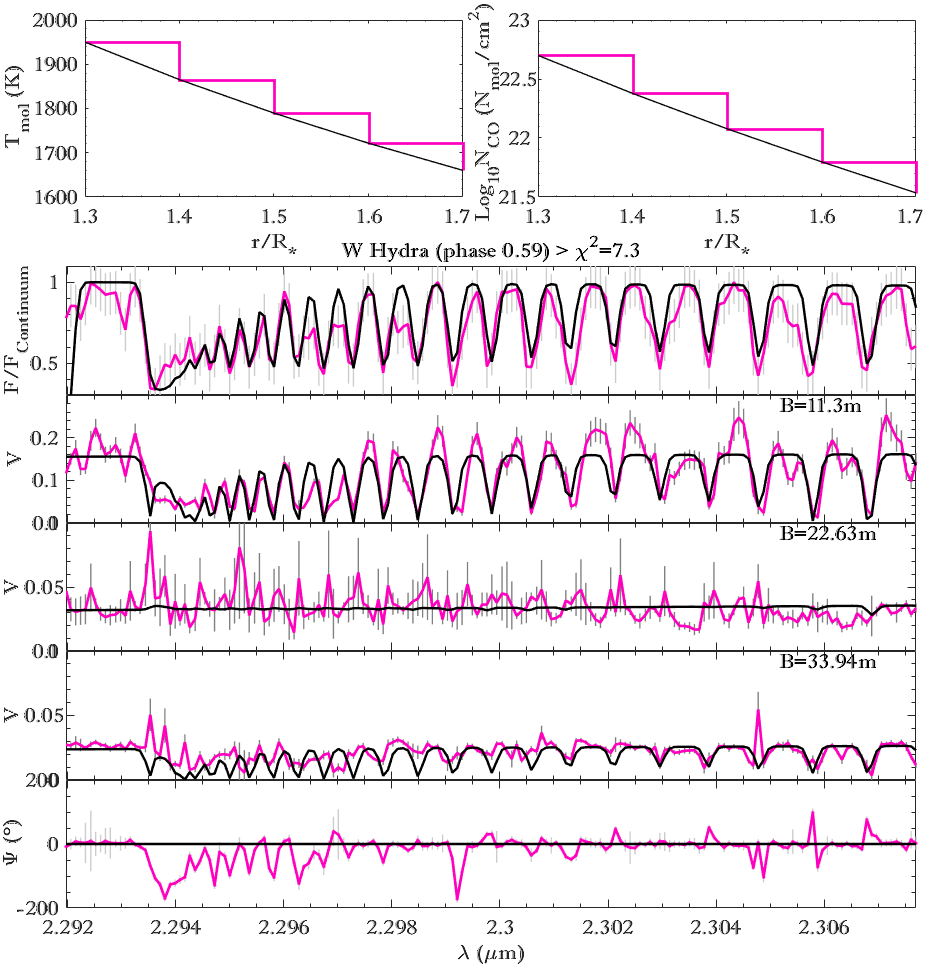}
\caption{Our best \textsc{PAMPERO} model for W Hya (phase 0.59) in the same manner as the Fig.~\ref{BKVir-Res}.}\label{whya-Res}
\end{figure}

\begin{figure}
\centering
\includegraphics[width=1.0\hsize,draft=false]{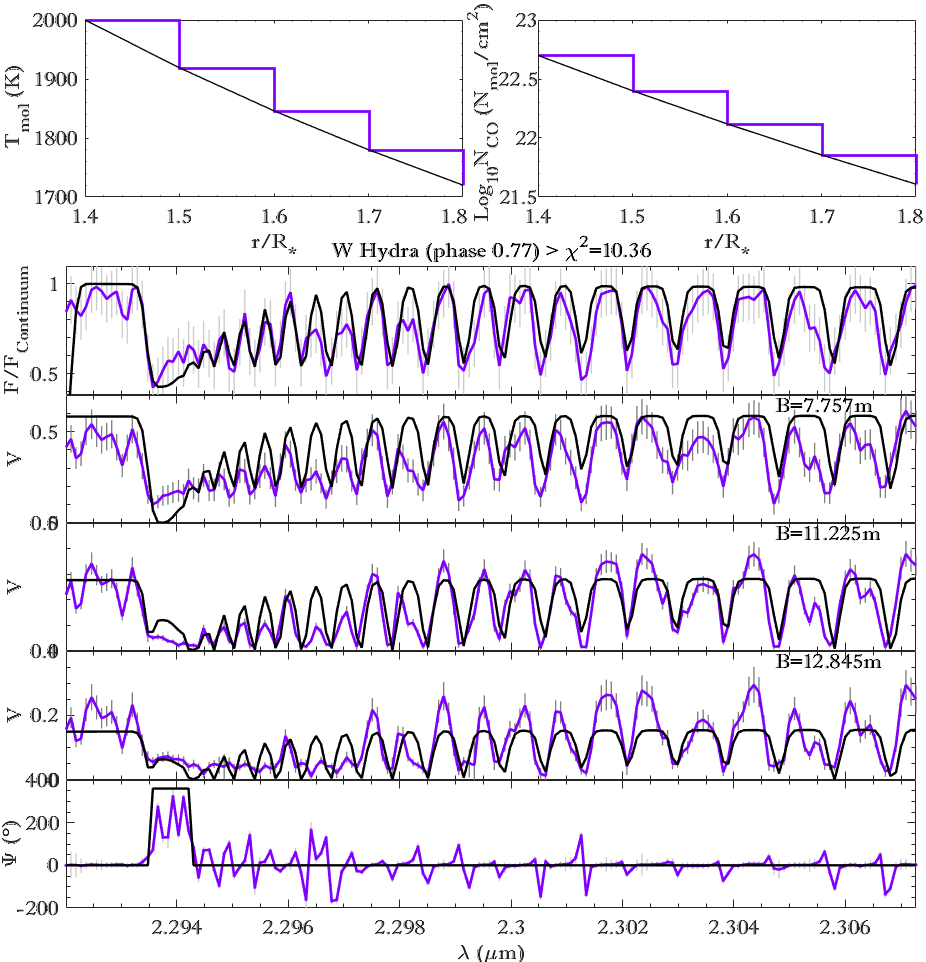}
\caption{Our best \textsc{PAMPERO} model for W Hya (phase 0.77) in the same manner as the Fig.~\ref{BKVir-Res}.}\label{whya2-Res}
\end{figure}

\section*{ACKNOWLEDGEMENTS }
\label{acknowledgments}
      This research made use of the SIMBAD database, operated at the CDS, Strasbourg, France, and of the NASA Astrophysics Data System Abstract Service. The author, M. Hadjara, acknowledges support from the scientific French association PSTJ \footnote{\url{http://www.pstj.fr/}} for its official host agreement, the Lagrange and OCA for computer server support. This research made use of the Jean-Marie Mariotti Center \texttt{SearchCal} service \footnote{Available at \url{http://www.jmmc.fr/searchcal}} codeveloped by Lagrange and IPAG, and of the CDS Astronomical Databases SIMBAD and VIZIER \footnote{Available at \url{http://cdsweb.u-strasbg.fr/}}. This research made use of the \texttt{AMBER data reduction package} of the Jean-Marie Mariotti Center\footnote{Available at \url{http://www.jmmc.fr/amberdrs}}. Special thanks go to the project's grant ALMA-CONICYT N$^\circ$ 31150002 and the PI, Keiichi Ohnaka, whose work inspired our new approach of the present work. Special thanks go to the project's grant ESO-MIXTO 2019, project's grant QUIMAL N$^\circ$ 3170082, as well as the grants from the Fizeau European interferometry initiative (I2E). Finally, MH and CN would like to express their warm thanks to Jeremy Tregloan-Reed (University of Antofagasta, Chile) for the precious help he provided for the correction of the English text of sections 5 and 6.


\end{document}